\newcommand{\be}{\begin{equation}}
\newcommand{\ee}{\end{equation}}
\newcommand{\bea}{\begin{eqnarray}}
\newcommand{\eea}{\end{eqnarray}}

\documentclass[a4paper,11pt,amsmath,amssymb,amsfonts,onecolumn,nofootinbib]{article}
\pdfoutput=1
\usepackage{jheppub}
\usepackage[T1]{fontenc}
\usepackage{graphicx}
\usepackage{dcolumn}
\usepackage{bm}
\usepackage{amsmath}
\usepackage{amssymb}
\def\d{d\kern-.8 ex\vrule height 1.3 ex depth-1.24 ex width .7 ex \kern .15 ex}
\def\D{D\kern-1.7 ex\vrule height .87 ex depth-.8 ex width .7 ex \kern .95 ex}

\title{Detecting few-body quantum chaos: out-of-time ordered correlators at saturation}

\author[a,b]{Dragan Markovi\'c}
\author[b]{and Mihailo \v{C}ubrovi\'c}
\affiliation[a]{Department of Physics, University of Belgrade, Studentski Trg 12-16, 11000 Belgrade, Serbia}
\affiliation[b]{Center for the Study of Complex Systems, Institute of Physics Belgrade, University of Belgrade, Pregrevica 118, 11080 Belgrade, Serbia}
\emailAdd{vokramnagard@gmail.com}
\emailAdd{cubrovic@ipb.ac.rs}

\date{\today}

\abstract{
We study numerically and analytically the time dependence and saturation of out-of-time ordered correlators (OTOC) in chaotic few-body quantum-mechanical systems: quantum Henon-Heiles system (weakly chaotic), BMN matrix quantum mechanics (strongly chaotic)
and Gaussian random matrix ensembles. The growth pattern of quantum-mechanical OTOC is complex and nonuniversal, with no clear exponential regime at relevant timescales in any of the examples studied (which is not in contradiction to the exponential growth
found in the literature for many-body systems, i.e. fields). On the other hand, the plateau (saturated) value of OTOC reached at long times decreases with temperature in a simple and universal way: $\exp(\mathrm{const.}/T^2)$ for strong
chaos (including random matrices) and $\exp(\mathrm{const.}/T)$ for weak chaos. For small matrices and sufficiently complex operators, there is also another, high-temperature regime where the saturated OTOC grows with temperature. Therefore, the plateau
OTOC value is a meaningful indicator of few-body quantum chaos. We also discuss some general consequences of our findings for the AdS/CFT duality. 

}

\begin{document}
\maketitle
\flushbottom

\section{Introduction}\label{secintro}

Recent years have seen a renewed interest in classical and quantum chaos in the context of high-energy physics, black holes and AdS/CFT, thanks to the relation of chaos to quantum information theory and the information problems of black holes.
Sharp and reasonably rigorous results such as the celebrated MSS chaos bound \cite{chaosbnd} and its subsequent refinements \cite{scramblestrings,chaosbnds} establish a connection between chaos and the fundamental properties of gravity and black
holes \cite{scramble,scramble2}. Maximal chaos, with the Lyapunov exponent $\lambda=2\pi T$ at temperature $T$, is reached for strongly coupled field theories in the large $N$ limit, which have a classical gravity dual with a black hole. In
\cite{butterstring} and other works it is explicitly shown how the Lyapunov exponent changes with finite $N$ effects.

However, it has been pointed out many times, also in the pioneering MSS paper \cite{chaosbnd}, that the multiple notions of quantum chaos in the literature mean different things. The out-of-time ordered correlation function (OTOC), given by the expectation
value of the commutator of some operators $A$ and $B$ at times $0$ and $t$:
\be
\label{otocdef0}C(t)\equiv\langle\vert\left[A(t),B(0)\right]\vert^2\rangle,
\ee
is a natural quantity in quantum field theories, i.e. many-body systems, and defines the quantum Lyapunov exponent $\lambda$ as the exponent of the time growth of OTOC. However, the classical limit of this exponent does not necessarily have much to do with
the classical Lyapunov exponent $\lambda_\mathrm{class}$, obtained by solving the variational equations \cite{lyap,lyapcao,hashimoto17}. The reason is the noncommutation of the three limits to be taken: the classical limit $\hbar\to 0$, the long-time limit
$t\to\infty$, and the small initial variation limit $\delta x(0)\to 0$. The crucial insight of \cite{lyapcao,hashimoto17} is that the mechanism of scrambling may be the chaotic dynamics, in which case $\lambda_\mathrm{class}$ is related though still not
identical to the OTOC exponent (quantum Lyapunov exponent) $\lambda$, or it may originate in local instability (hyperbolicity), in which case even regular systems may have a nonzero $\lambda$ exponent and likewise chaotic systems may have $\lambda$ which
is completely  unrelated to the classical counterpart. 

This mismatch between the classical and quantum Lyapunov exponent is just the tip of the iceberg. The problem is twofold: not only what is the relation between the quantum (OTOC) exponent and classical chaos, but also what is the relation between the 
quantum Lyapunov exponent $\lambda$ and other indicators of quantum chaos such as, first and foremost, level statistics. The bread and butter of quantum chaos is the famous Dyson threefold way leading to the Wigner surmise, the level repulsion statistics
determined solely by the time reversal properties of the Hamiltonian \cite{thebook}, which follows from the random matrix approximation of chaotic Hamiltonian operators \cite{randmat}. It is no secret for several years already that the black hole
quasinormal mode spectra follow the random matrix statistics \cite{cotlerbig}, and the OTOC of a Gaussian unitary ensemble (GUE) has been computed analytically in terms of Bessel functions in \cite{cotlerotoc,cotlerdecoup}; the outcome is close to the
expected behavior of large-$N$ field theories only at long timescales, longer than the scrambling time; at shorter timescales there are important differences. The authors of \cite{cotlerotoc} have reached a deep conclusion in this respect: random
matrices have no notion of locality as the correlation of any pair of eigenvalues is described by the same universal function. This is why the OTOC of a GUE system deviates from that of a local field theory at early times, when the perturbation in
field theory has not had time to spread yet (i.e. when it is still localized). Therefore, the level repulsion does not imply the usual picture of the chaotic (exponential) OTOC behavior. However, we do not know yet how this correlates to the behavior of
\emph{few-body} or more precisely few-degrees-of-freedom quantum systems as opposed to the large-$N$ field theories with a gravity dual. In few-body systems the notion of locality (and a classical gravity dual) does not exist anyway and the main problem
found by \cite{cotlerotoc} is irrelevant; at the same time, such systems are often very well described by random matrix statistics, i.e. Wigner-Dyson statistics \cite{thebook}. In this paper we aim to understand the behavior of OTOC in such systems.
Running a bit forward, we can say that the \emph{growth} of OTOC is rather unremarkable: we find no universal trend, and little connection to level statistics. This confirms the results found for specific examples in \cite{hashimoto17,false}.

The relation of OTOC, level statistics and the classical Lyapunov exponent was studied for few-body systems (quantum mechanics) in \cite{hashimoto17,false,inverse,hashimoto20a,poljaci} and the picture is inconclusive. One can
have a nonzero growth exponent in integrable systems,\footnote{This actually correlates with the classical variational equations in hyperbolic systems, which show exponential growth even in absence of chaos.} whereas fast scrambling with the exponent close
to $2\pi T$ or at least growing linearly in $T$ has not been found even in some clearly chaotic systems \cite{hashimoto17,hashimoto20b}. Arguments for many-body systems such as spin chains even suggest that quantum-chaotic systems with Gaussian spectral
statistics generically never show fast scrambling \cite{prosenspin}, but no claims of such generality have been tested or formulated for few-body quantum chaos.

Various indicators of chaos relevant also for small systems, and their relation to OTOC and scrambling were studied by \cite{pappal,okuyama,richter,kerr,doublekicked,rabi,kickedhussimi,energylimit} among others. In particular, in \cite{kickedhussimi} some
important  insights can be found: even in small systems devoid of the locality notion, OTOC can be interpreted as a measure of delocalization of a state in phase space, and the oscillatory component of the OTOC dynamics has to do with the power spectrum of
the system. This last insight provokes a more general question: can we learn something from the quasi-stationary regime of OTOC, where no systematic growth is present but only oscillations? In this paper we provide a partial answer from a detailed study
of this saturated (asymptotic, plateau) OTOC regime: the magnitude of the OTOC average at the plateau has a simple temperature dependence, and apparently can differentiate between weak chaos (dominantly Poissonian level distribution with some admixture of
the Wigner-Dyson statistics) and strong chaos (clear Wigner-Dyson level repulsion). We will demonstrate this on three representative systems: the quantum Henon-Heiles Hamiltonian, whose classical limit has mixed (regular/chaotic) phase space and thus we
expect on average weak chaos, a simplified BMN matrix model (at small $N$) exhibiting strong chaos for most initial conditions, and Gaussian random matrices, the prototype of strong quantum chaos. The long-time limit of OTOC behaves in subtly different
ways in each case.

Before we start, one caveat is in order (we will consider this issue in more detail later on): one might think that the saturated OTOC value is always trivially determined by the system size. We typically assume that the OTOC function $C$ as
defined in (\ref{otocdef0}) behaves roughly as $C(t)\sim c/N^2\times\exp(\lambda t)$ with $c$ of order unity, so when $t\sim t_*\equiv\log N^2/\lambda$ the growth of $C(t)$ stops and OTOC approximately reaches unity (when
appropriately normalized). But the twist is precisely that $c$ is system-specific and in general poorly known. The leading $N^2$ behavior indeed determines the OTOC values for $N$ large, but when $N$ and $c$ are comparable within an order of magnitude the
effects of fluctuations and finite $N$ corrections are significant. This is at the root of our observations in this work.


The plan of the paper is as follows. In section \ref{secotocqm} we recapitulate and generalize some results on computing OTOC in quantum mechanics, and show how OTOC sensitively depends on both the Hamiltonian and the operators $A$, $B$ from the
definition (\ref{otocdef0}). In section \ref{secotocrm} we apply the general formalism to random matrix ensembles and show that the OTOC growth is a complicated and nonuniversal function but that its asymptotic value behaves in a rather simple way. Section
\ref{secotocham} discusses the behavior of OTOC for deterministic quantum-chaotic Hamiltonians. Section \ref{secconc} sums up the conclusions.

\section{OTOC in quantum-mechanical systems}\label{secotocqm}

Consider a four-point time-disordered correlation function for a quantum-mechanical system in $0+1$ dimensions at temperature $T=1/\beta$. Starting from the usual definition (\ref{otocdef0}) as the squared module of the commutator of the two operators $A$
and $B$, we can write it out as
\be
\label{otocdef}C(t)=\frac{1}{Z}\langle\vert\left[A(t),B(0)\right]\vert^2\rangle=\frac{1}{Z}\sum_ne^{-\beta E_n}\langle n\vert\vert\left[A(t),B(0)\right]\vert^2\vert n\rangle,
\ee
where the averaging is both thermal and quantum mechanical: $\langle\ldots\rangle=\mathrm{tr}e^{-\beta H}\langle\mathrm{vac}\vert\ldots\vert\mathrm{vac}\rangle$. We can pick a basis of states and express the above defining expression in terms of matrix
elements of the operators (this closely follows the derivation in \cite{hashimoto17,hashimoto20a}):
\be
\label{otocdef1}C(t)=\frac{1}{Z}\sum_{nm}e^{-\beta E_n}\langle n\vert\left[A(t),B(0)\right]\vert m\rangle\langle m\vert \left[A(t),B(0)\right]\vert n\rangle=\frac{1}{Z}\sum_{nm}e^{-\beta E_n}\vert c_{mn}(t)\vert^2,
\ee
where we have inserted the completeness relation $1=\sum_m\vert m\rangle\langle m\vert$. For a single element $c_{mn}(t)$ one gets:
\bea
\nonumber &&c_{mn}(t)=\langle n\vert\left[e^{\imath Ht}Ae^{-\imath Ht},B\right]\vert m\rangle=\\
\nonumber&&=\sum_k\bigg(\langle n\vert e^{\imath Ht}Ae^{-\imath Ht}\vert k\rangle\langle k\vert B\vert m\rangle-\langle n\vert B\vert k\rangle\langle k\vert e^{\imath Ht}Ae^{-\imath Ht}\vert m\rangle\bigg)=\\
\nonumber&&=\sum_k\bigg(\langle n\vert e^{\imath E_nt}Ae^{-\imath E_kt}\vert k\rangle\langle k\vert B\vert m\rangle-\langle n\vert B\vert k\rangle\langle k\vert e^{\imath E_kt}Ae^{-\imath E_mt}\vert m\rangle\bigg)=\\
\label{otocmn}&&=\sum_k\left(a_{nk}b_{km}e^{-\imath E_{kn}t}-b_{nk}a_{km}e^{-\imath E_{mk}t}\right),
\eea
where in the second line we have again inserted a completeness relation and in the third line we have used the fact that we work in the energy eigenbasis. The outcome is expressed in terms of the matrix elements $a_{mn},b_{mn}$ of the operators in the
energy basis. In practice, it may or may not be possible to compute these analytically. Specifically, for $A=x,B=p$, we get the analogue of the classical Lyapunov exponent. From now on we call this the kinematic OTOC as it is directly related to the
classical trajectory. Let us now see what general bounds can be put on (\ref{otocmn}) from the properties of quantum-mechanical Hamiltonians.

\subsection{An upper bound on OTOC saturation}

We begin with a very general and very formal result, which immediately makes it clear that in a generic quantum-mechanical system (integrable or nonintegrable) OTOC can be bounded from above by a quantity which solely depends on the energy spectrum of the 
Hamiltonian and the choice of the operators $A$ and $B$. This upper bound remains valid no matter what is the time dependence of OTOC, even if it does not have a nonzero growth exponent at all (which is quite generic in quantum mechanics). Starting from
the basic equations (\ref{otocdef1}-\ref{otocmn}), let us denote $C_{nmk}=a_{nk}b_{km}$ i $D_{nmk}=-b_{nk}a_{km}$, and estimate a single coefficient $c_{mn}(t)$ in the sum. We clearly have 
\bea
\nonumber\vert c_{mn}(t)\vert&=&\vert\sum_kC_{nmk}e^{-iE_{kn}t}+D_{nmk}e^{-iE_{mk}t}\vert\leq\sum_k\vert C_{nmk}e^{-iE_{kn}t}+D_{nmk}e^{-iE_{mk}t}\vert\Rightarrow\\
\nonumber\vert c_{mn}(t)\vert^2&\leq&\bigg(\sum_k\vert C_{nmk}e^{-iE_{kn}t}+D_{nmk}e^{-iE_{mk}t}\vert\bigg)^2\leq \sum_k\vert C_{nmk}e^{-iE_{kn}t}+D_{nmk}e^{-iE_{mk}t}\vert^2\leq\\
&\leq&\sum_k\left(\vert C_{nmk}\vert^2+\vert D_{nmk}\vert^2+2\vert C_{nmk}\vert\vert D_{nmk}\vert\cos\left(E_{mk}-E_{kn}\right)t\right),
\eea
where $N$ is the matrix size. In the second and third line we have used the inequality between the arithmetic and harmonic mean. Now we can bound the value of $C(t)$:
\be
\label{otocineq0}0\leq C(t)\leq\frac{1}{Z}\sum_{nmk}e^{-\beta E_n}\left(\vert C_{nmk}\vert^2+\vert D_{nmk}\vert^2+2\vert C_{nmk}\vert\vert D_{nmk}\vert\cos\left(E_{mk}-E_{kn}\right)t\right)
\ee
This means that $C(t)$ is bounded at all times by an oscillatory function of time, whose frequencies are linear combinations of three eigenenergies ($E_{mk}-E_{kn}=E_m+E_n-2E_k$). Such a combination is generically always nonzero for a chaotic system except
when the energies coincide, e.g. $E_m=E_n=E_k$ (according to the non-resonance condition). Therefore, since OTOC is typically a non-decreasing function of time, the behavior of $C(t)$ for $t$ large is roughly its maximum value and is likely close to the
right-hand side in (\ref{otocineq0}). This suggests that the OTOC dynamics after saturation likely consists of a very complex oscillatory pattern (with $\sim N^3$ frequencies if the Hilbert space has dimension $N$) superimposed on a plateau. The numerics
will indeed confirm such behavior.

Another estimate, which is time-independent and relevant for our main result -- the magnitude of the saturation (plateau) OTOC value, is obtained from the triangle and mean inequalities:
\bea
\nonumber\vert c_{mn}(t)\vert^2&\leq&\vert\sum_kC_{nmk}e^{-iE_{kn}t}\vert^2+\vert\sum_kD_{nmk}e^{-iE_{mk}t}\vert^2\leq\vert\sum_kC_{nmk}\vert^2+\vert\sum_kD_{nmk}\vert^2\Rightarrow\\
\label{otocineq}C(t)&\leq&\frac{1}{Z}\sum_{nm}e^{-\beta E_n}\left(\vert\left(A\cdot B\right)_{nm}\vert^2+\vert\left(B\cdot A\right)_{nm}\vert^2\right)\leq\frac{2}{Z}\sum_{nm}e^{-\beta E_n}\vert\left(A\cdot B\right)_{nm}\vert^2,~~~~~~~
\eea
where we have used the obvious relations $\sum_kC_{mnk}=(A\cdot B)_{nm}$ and $\sum_kD_{nmk}=(B\cdot A)_{nm}=(A\cdot B)^*_{mn}=(A\cdot B)_{nm}$, assuming also the hermiticity of the operators. For some models (e.g. random matrices, Henon-Heiles), this sum 
can be estimated in a controlled way and provides an approximation for the plateau of OTOC. These estimates are obviously very simple and very weak (in the mathematical sense) but provide us with a framework into which we can insert specific $A$, $B$ and 
$H$ (the Hamiltonian with energies $E_n$) and perform back-of-the-envelope calculations which explain the numerical findings.

\section{OTOC for random matrix ensembles}\label{secotocrm}


Random matrix theory \cite{randmat,thebook} provides a highly detailed and rigorous (within its starting assumptions) stochastic effective description of the few-body quantum chaos, and allows an analytic calculation of OTOC along
the lines of (\ref{otocmn}). Let us focus on Gaussian ortohogonal ensembles of size $N\times N$, appropriate when there is full time reversal invariance. It is known \cite{thebook} that the joint distribution all the elements of all
eigenvectors is obtained simply from the statistical independence of the eigenvectors from each other and of the elements in each eigenvector (and the orthogonality of the eigenvectors):
\be
P\left(\lbrace c\rbrace\right)=\left(\prod_{n=1}^N\delta\left(\sum_i\left(c^n_i\right)^2-1\right)\right)\left(\prod_{n<m}\delta\left(\sum_ic^n_ic^m_i\right)\right),\label{randmatp}
\ee
where $i=1\ldots N$ is the component of the eigenvector and $1\leq n,m\leq N$ count the eigenvectors themselves, i.e. the energy levels; so the $n$-th eigenvector $\vert n\rangle$ is represented by the column vector $\psi^{(n)}$ with the elements
$(c^n_1,\ldots c^n_N)$. Special cases like the probability distribution for the $p$-tuple of elements of a single eigenvector are obtained from (\ref{randmatp}) by integrating out all the other elements \cite{thebook}. We will also need the probability
distribution of the energy levels $\lbrace E\rbrace={E_1,E_2,\ldots E_N}$, the celebrated Wigner-Dyson distribution function \cite{thebook}:
\be
\mathcal{P}\left(\lbrace E\rbrace\right)=\mathrm{const.}\times\prod_{n<m}\vert E_n-E_m\vert^b\exp\left(-\sum_k\frac{E_k^2}{\sigma^2}\right),
\ee
where $\sigma$ is the standard deviation, fixing the unit of energy, and $b=1$, $2$ or $4$ for orthogonal, unitary and symplectic ensembles respectively. Most of our work is independent of the symmetry class, however our default class will be the Gaussian 
orthogonal ensemble (GOE) with $b=1$ when not specified otherwise.

\subsection{Estimate of the OTOC and its plateau}

The idea is to use the results recapitulated in the previous section to find the ensemble expectation value of OTOC from the "master formulas" (\ref{otocdef1}-\ref{otocmn}). Representing the eigenvectors and the operators as matrices in some (arbitrary)
basis we can obviously write out
\be
a_{nk}=\sum_{ij}\psi^{(n)}_i\psi^{(k)}_jA_{ij}\Rightarrow\langle a_{nk}\rangle=\int d^N\psi^{(n)}\int d^N\psi^{(k)}P\left(\psi^{(n)},\psi^{(k)}\right)\psi^{(n)}_i\psi^{(k)}_jA_{ij},
\ee
and similarly for $b_{nk}$. Inserting the above expression for the matrix elements into (\ref{otocmn}), multiplying $c_{mn}(t)$ by its complex conjugate taking into account the reality of the eigenvectors and relabelling the indices in the sums where
convenient we find (denoting the average over the random matrix ensemble by $\langle C(t)\rangle$):
\bea
\nonumber  \langle C(t)\rangle&=&\int d^{N^2}\lbrace c\rbrace\int d^N\lbrace E\rbrace\mathcal{P}\left(\lbrace E\rbrace\right)P\left(\lbrace c\rbrace\right)
\sum_{n,m}\sum_{k,k'}\sum_{i_{1,2}}\sum_{j_{1,2}}\sum_{i'_{1,2}}\sum_{j'_{1,2}}c_{j_1}^kc_{i_2}^kc_{j'_1}^{k'}c_{i'_2}^{k'}c^n_{i_1}c^n_{i'_1}c^m_{j_2}c^m_{j'_2}e^{-\beta E_n}\times\\
\nonumber &\times&\bigg(A_{i_1i_2}A_{i'_1i'_2}B_{j_1j_2}B_{j'_1j'_2}e^{\imath(E_{k'}-E_k)t}+A_{i_2j_2}A_{i'_2j'_2}B_{i_1j_1}B_{i'_1j'_1}e^{\imath(E_k-E_{k'})t}-\\
&-&A_{i_2j_2}A_{i'_1i'_2}B_{i_1j_1}B_{j'_1j'_2}e^{\imath(E_k+E_{k'}-E_m-E_n)t}-A_{i_1i_2}A_{i'_2j'_2}B_{j_1j_2}B_{i'_1j'_1}e^{\imath(E_m+E_n-E_k-E_{k'})t}\bigg),\label{otocmaster}
\eea
where $\lbrace c\rbrace$ determines the whole set of $N^2$ random elements $c^{(n)}_j$ with $j,n=1\ldots N$ and likewise $\lbrace E_n\rbrace$ is the whole set of eigenenergies. All the sums run from $1$ to $N$. The integral over the eigenvector elements
$\lbrace c\rbrace$ in (\ref{otocmaster}) produces only an overall constant as these coefficients do not couple to the other quantities (in fact the integral $d^{N^2}\lbrace c\rbrace$ is a textbook Jeans integral, but we do not need its value as it only
produces an $N$-dependent, $T$-independent constant). The remaining integral, over the eigenenergies, is again a sum of products of Jeans-type integrals but with an additional linear term $-\beta E$ in the exponent. Notice that the imaginary (sine) terms
in (\ref{otocmaster}) cancel out when the sum is performed; this is a consequence of the module squared in $\vert c_{mn}\vert^2$, i.e. of the reality of OTOC. Now we see that (\ref{otocmaster}) becomes a sum where each term is a product of factors of the
form
\be
p_i(E_i)e^{-E_i^2/4\sigma^2-\beta E_i}\cos(sE_it),~~s\in\lbrace 0,1\rbrace,\nonumber
\ee
where $p_i$ is some polynomial and $s$ may be zero or unity, i.e. some terms have this factor and some do not. Every such term is a Jeans-type integral. The number of terms in $\mathcal{P}\left(\lbrace E\rbrace\right)$ equals the number of partitions of
$N(N-1)b/2$, and the sums over the coefficients $\lbrace c\rbrace$ bring alltogether $N^{12}$ terms. When everything is said and done (for details see Appendix \ref{secappgoe}), the final outcome, ignoring the multiplicative constant factors, reads: 
\bea
\nonumber \langle C(t)\rangle&=&\prod_{a=1}^4\sum_{\alpha^a_1,\ldots\alpha^a_N}^{\sum_j\alpha^a_j=N(N-1)b/2}
\bigg[{}_1F_1\left(\frac{1+\alpha^a}{2},\frac{1}{2},\frac{\sigma^2}{4}\left(\beta-\imath t\right)^2\right)+(\beta-\imath t){}_1F_1\left(\frac{2+\alpha^a}{2},\frac{3}{2},\frac{\sigma^2}{4}\left(\beta-\imath t\right)^2\right)+\\
&+&{}_1F_1\left(\frac{1+\alpha^a}{2},\frac{1}{2},\frac{\sigma^2}{4}\left(\beta+\imath t\right)^2\right)+(\beta+\imath t){}_1F_1\left(\frac{2+\alpha^a}{2},\frac{3}{2},\frac{\sigma^2}{4}\left(\beta+\imath t\right)^2\right)\bigg],\label{otocrandtemp}
\eea
where ${}_1F_1$ is the confluent hypergeometric function. The sum runs over all partitions of $N(N-1)b/2$, and the product has four terms as each factor $\vert c_{mn}\vert^2$ has four matrix elements of $A$ and $B$.

\subsubsection{Kinematic OTOC}

In order to move further we need to specify at least to some extent the operators $A$ and $B$. We will consider (1) the kinematic OTOC, with $A=x$, $B=p$ (2) generic sparse operators, with $O(N)$ nonzero elements in the matrices $a_{mn}$ and $b_{mn}$, and
(3) dense operators $A$ and $B$, with $O(N^2)$ nonzero elements, in particular the case when the operators $A$, $B$ are themselves represented by Gaussian random matrices. Let us estimate OTOC for each case.

For the kinematic OTOC, $A_{ij}=x_i\delta_{ij}$ is diagonal and in the large-$N$ limit $B$ can be approximated as $B_{ij}\sim\delta_{ij}/x_i$. The Kronecker deltas reduce the number of terms in the sums over $\lbrace c\rbrace$ to $N^4$, the number of
partitions $\sum_j\alpha_j=n$ can be approximated as $p(n)\sim\exp(\pi\sqrt{2n/3})/\sqrt{n}$, and the general expression (\ref{otocmaster}) becomes\footnote{One might be surprised by the unusual dependence on $N$. This happens because we have not
normalized $C(t)$ by the product $\langle AA\rangle\langle AA\rangle$ as it is usually done. With appropriate normalization, $C(t)$ would of course be of order unity.}
\be
\langle C(t)\rangle\sim e^{\pi\sqrt{\frac{b}{3}}N}N^3e^{\frac{\sigma^2\beta^2}{4}}
\left(W_0\left(\sigma\beta\right)+Q_1\left(\cos\frac{\sigma^2\beta t}{2}\right)W_1\left(\sigma\beta\right)+Q_2\left(\sin\frac{\sigma^2\beta t}{2}\right)W_2\left(\sigma\beta\right)\right)\label{otocrandxp},\\
\ee
where $W_{0,1,2}$ are polynomials in $\sigma\beta$ of degree $N(N-1)b/2\sim N^2b/2$, $Q_1$ is an even polynomial (with only even powers) of the same degree, and $Q_2$ is an odd polynomial of the same degree. Each coefficient 
in the polynomials $W_{0,1,2}$ comes from $\sim N^2$ terms (Appendix \ref{secappgoe}), therefore the size of the coefficients scales approximately with $N^2$. Eq.~(\ref{otocrandxp}) is a very complicated oscillating function as many terms are involved. But
if we are only interested in the average value of $C(t)$ at long times, we may simply ignore the oscillations (which in the first approximation average out to some value of order unity) and write the
estimate for the long-term, saturated or plateau OTOC value that we denote by $C_\infty$:
\be
\label{otocrandxpinf}C_\infty\sim\langle C(t\to\infty)\rangle\sim e^{\pi\sqrt{\frac{b}{3}}N}N^3e^{\frac{\sigma^2\beta^2}{4}}W_0\left(\sigma\beta\right)
\ee
We deliberately do not write $\lim_{t\to\infty}$ in the above definition as the limit in the strict sense does not exist because of the oscillatory functions, and in addition our derivation is obviously nothing but a crude estimate. A similarly rough
estimate of the temperature dependence of $C_\infty$ can be obtained in the following way. For sufficiently large $\sigma\beta$, roughly $\sigma\beta/N^2>1$, the polynomial $W_0$ is dominated by the highest-degree term and we have, from
(\ref{otocrandxpinf}):
\be
\label{otocrandxplarge}C_\infty\sim\left(\sigma\beta\right)^{\frac{N^2b}{2}}e^{\frac{\sigma^2\beta^2}{4}}+\ldots\sim e^{\frac{\sigma^2}{4T^2}}+\ldots,
\ee
where in the second step we have assumed $\beta\gg 1$ so that the power-law prefactor $\beta^{N^2b/2}$ becomes negligible compared to the exponential. We deliberately emphasize that there are other terms in the expansion ($\ldots$), including also a
constant term (from the zeroth-order term in $W_0$). This is important as it tells us that the scaling is in general of the form $C_\infty\approx\mathrm{const.}+\exp(\sigma^2/4T^2)$, i.e. the temperature dependence is superimposed to a constant. This is
also expected as the (appropriately normalized) saturated value $C_\infty$ should always be of order unity, and the temperature dependence will only account for the relatively small differences between the plateau values of $C(t)$, as we will see later in
Figs.~\ref{fignumotoc1} and \ref{fignumotoc2}.

On the other hand, for sufficiently small $\sigma\beta$, the polynomial $W_0$ can be estimated as a geometric sum of monomials in $-\sigma\beta N^2$ (remember the terms in $W_0$ have alternating signs):
\be
\label{otocrandxpsmall}C_\infty\sim\frac{e^{\frac{\sigma^2\beta^2}{4}}}{1+\sigma\beta N^2}\sim 1-\sigma\beta N^2+O\left(\beta^2\right).
\ee
We have now reached an important point: the plateau OTOC falls off exponentially with $1/T^2$ at low temperatures\footnote{Actually, the falloff rate equals $\mathrm{const.}/T^2$ with some system-specific constant, but for brevity we will denote it
schematically as the $1/T^2$ regime throughout the paper.} and grows as a function of $1/T$ at high temperatures (we are not sure which function, as there are higher order terms in addition to the one written in (\ref{otocrandxpsmall}), and there is no
clearly dominant term like the exponential at large $\beta$), with the crossover temperature:\footnote{The crossover temperature is determined simply as $\beta_c\sigma N^2=1$, i.e. whether the terms in $W_0$ grow or decay at higher and higher order.} 
\be
\label{tcrossrandxp}T_c\sim\sigma N^2.
\ee
If we consider a pair of arbitrary sparse operators $A$ and $B$, the whole above reasoning remains in place, except that the products of matrix elements such as $A_{i_1i_2}A_{i'_1i'_2}B_{j_1j_2}B_{j'_1j'_2}$ remain as arbitrary constants. Therefore we get
the same qualitative behavior with two regimes and a crossover between them. The crossover temperature is very high for typical $N\gg 1$ (otherwise the random matrix formalism makes little sense) and finite $\sigma$ (again, $\sigma\to 0$ makes little
sense). In particular, in the $N\to\infty$ limit the crossover temperature becomes infinite and the only regime is the exponential decay.

\subsubsection{OTOC for dense and/or random operators}

Now consider the case when the matrix elements in (\ref{otocrandtemp}) are generically all nonzero (and for now nonrandom, i.e. we fix the operators and do not average over them). The large-$t$ limit yields the expression
\be
\label{otocranddense}\langle C(t)\rangle\sim e^{\pi\sqrt{\frac{b}{3}}N}N^{11}e^{\sigma^2\left(\beta^2-t^2\right)}
\left[q_0\left(\sigma t\right)w_0\left(\sigma\beta\right)+q_1\left(\sigma t\right)w_1\left(\sigma\beta+\imath t,\sigma\beta-\imath t\right)\right],
\ee
where $q_{0,1}$ and $w_0$ are polynomials of degree $N(N-1)b/2\sim N^2b/2$, and $w_1$ is the polynomial of the same total degree of two variables, $\sigma\beta+\imath t$ and $\sigma\beta-\imath t$. The coefficients of $w_{0,1}$ are proportional to products
of matrix elements $A_{i_1j_1}B_{k_1l_1}\ldots A_{i_8j_8}B_{k_8l_8}$, which are roughly proportional to $\vert A\vert^8\vert B\vert^8$. The long-time limit yields
\be
\label{otocrandinf}C_\infty\sim\frac{e^{\pi\sqrt{\frac{b}{3}}N}}{N^5}\left(\vert A\vert\vert B\vert\right)^8e^{\frac{\sigma^2\beta^2}{4}}w_0(\sigma\beta)w_1(\sigma\beta,\sigma\beta),
\ee
but now a typical coefficient of the polynomials $w_{0,1}$ behaves as $N^2\left(\vert A\vert\vert B\vert\right)^8$. Therefore, the scaling in the low-temperature regime remains the same as (\ref{otocrandxplarge}): $C_\infty\sim\exp(\sigma^2/T^2)$. But the
high-temperature regime yields
\be
\label{otocrandsmall}C_\infty\sim 1-\sigma\beta N^2\left(\vert A\vert\vert B\vert\right)^8+O\left(\beta^2\right),
\ee
therefore the crossover now happens at 
\be
\label{tcrossrand}T_c\sim\sigma N^2\left(\vert A\vert\vert B\vert\right)^8
\ee
and therefore may be lower than the very high value (\ref{tcrossrandxp}), depending on the norm of the operators $A$ and $B$.

Finally, if the operators $A$ and $B$ are both random Hermitian matrices (for concreteness, from the Gaussian unitary ensemble with the distribution function $\Pi$ and the standard deviation $\xi$), we need to average also over the distribution functions
for $A$ and $B$ and work with the double average $\langle\langle C(t)\rangle\rangle$:
\be
\langle\langle C(t)\rangle\rangle\equiv\int d^N\lbrace a\rbrace\int d^N\lbrace b\rbrace\Pi\left(\lbrace a\rbrace\right)\Pi\left(\lbrace b\rbrace\right)\langle C(t)\rangle\sim\xi^{N^2}\langle C(t)\rangle.
\ee
This estimate is very crude, based simply on the fact that the distribution functions $\Pi$ have $\sim N^2/2$ pairs of the form $(a_i-a_j)^2$. The important point is that the scaling from (\ref{otocrandsmall}) that behaves essentially as
$\sim\xi^{16}$ now becomes $\sim\xi^{N^2}$, therefore the crossover temperature is significantly reduced compared to (\ref{tcrossrand}) and behaves as $T_c\sim\sigma N^2\lambda^{N^2}$. So for random operators the crossover may
happen at temperatures that are not very high and thus can be clearly visible in the numerics and experiment.

\subsection{Numerical checks}

\begin{figure}[ht]
(A)\includegraphics[width=0.4\textwidth]{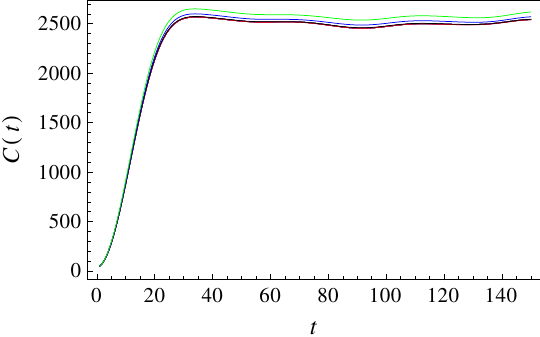}
(B)\includegraphics[width=0.4\textwidth]{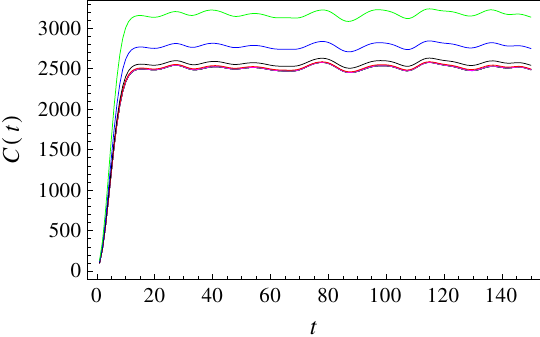}\\
(C)\includegraphics[width=0.4\textwidth]{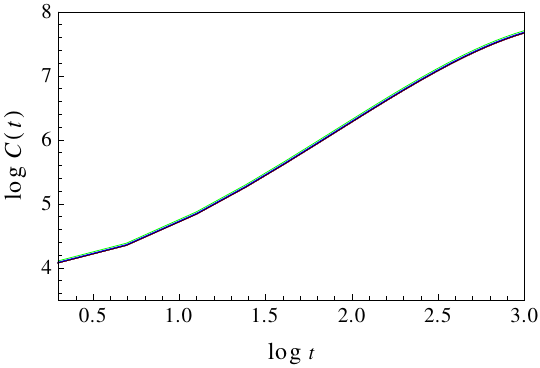}
(D)\includegraphics[width=0.4\textwidth]{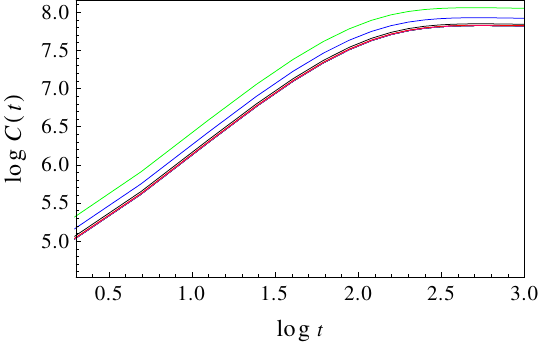}\\
\caption{\label{fignumotoc1} Numerically computed and averaged kinematic OTOC $C(t)$ for an ensemble of $l=1000$ Gaussian orthogonal matrices of size $N=20$ for the deviation $\sigma=0.02$ (A, C) and $\sigma=0.05$ (B, D), at temperatures $0.67$ (black),
$1.00$ (blue), $1.25$ (green), $2.50$ (magenta), $5$ (red). The plots (A, B) show the linear scale and the plots (C, D) the log-log scale. Crucially, the growth regime is not exponential and is actually closer to a power law. The growth ends on a plateau
with superimposed oscillations. The plateaus differ slightly for different temperatures -- the main effect we look at in this paper. Times is in units $1/\sigma$ in all plots.}
\end{figure}

\begin{figure}[ht]
(A)\includegraphics[width=0.4\textwidth]{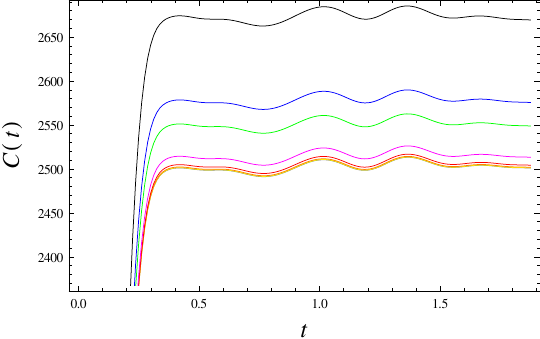}
(B)\includegraphics[width=0.4\textwidth]{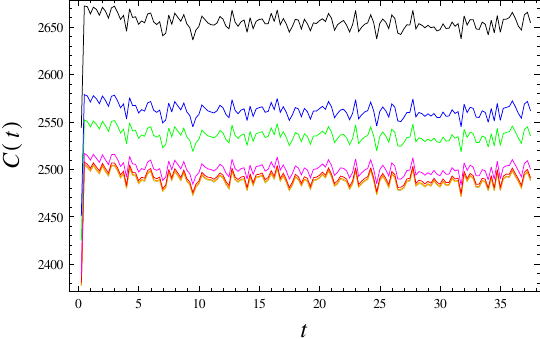}\\
\caption{\label{fignumotoc2} Numerically computed and averaged kinematic OTOC $C(t)$ for an ensemble of $l=1000$ Gaussian orthogonal matrices of size $N=60$ for the deviation $\sigma=0.1$, at temperatures $0.67$ (black), $1.00$ (blue), $1.25$ (green),
$2.50$ (magenta), $5$ (red), $10$ (orange), $20$ (yellow) and $100$ gray. In (B) we plot the same as in (A) but over a longer timescale, showing that the plateau remains stable for long times, i.e. represents true asymptotic behavior.}
\end{figure}

Now we demonstrate numerically that the crude estimates from the previous subsection indeed make sense and describe the characteristic behavior of OTOC. Our chief goal is to understand the behavior of $C_\infty$, however it is instructive to start from 
the time dependence of the kinematic OTOC (Fig.~\ref{fignumotoc1}). We find the expected pattern of early growth followed by a plateau, however the growth is closer to a power law than to an exponential; this follows from the polynomial terms in
(\ref{otocrandxp}), although the power law is not perfect either, as we see in the panels (A, C). This is in line with the prediction of \cite{cotlerotoc}, where the authors find
\be
\langle C(t)\rangle\approx J_1^4(2t)/t^4+t(t/2-1),\label{cotler}
\ee
for a slightly different ensemble of random matrices ($J_1$ is the Bessel function of the first kind). This function is also neither an exponential nor a power law but at early times it is best approximated by a power law at leading order (at long times it
falls off exponentially but the saturation is reached already prior to that epoch). In Fig.~\ref{fignumotoc2} we focus on the plateau behavior. It has the form of a constant function with superimposed aperiodic oscillations, and the differences of the
plateau values are the subject of our theoretical predictions. These are relatively small and become important only when $N$ is finite and not very large. In Fig.~\ref{fignumotoc2} we plot again the time dependence of the kinematic OTOC but now we focus on
long timescales, to confirm that the plateau is indeed stable, and to show the very complex oscillation pattern superimposed on the plateau.

\begin{figure}[ht]
(A)\includegraphics[width=0.28\textwidth]{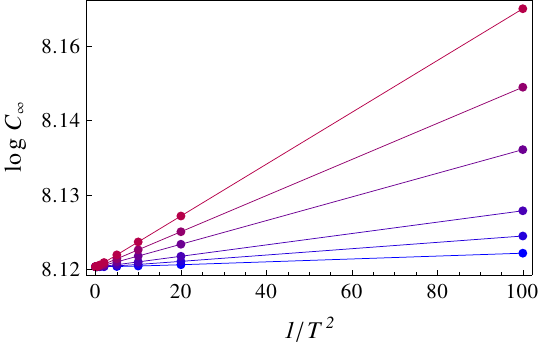}
(B)\includegraphics[width=0.28\textwidth]{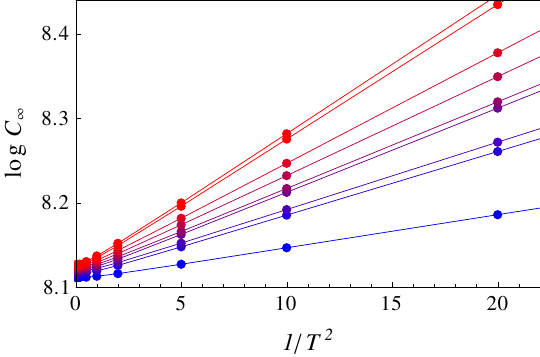}
(C)\includegraphics[width=0.28\textwidth]{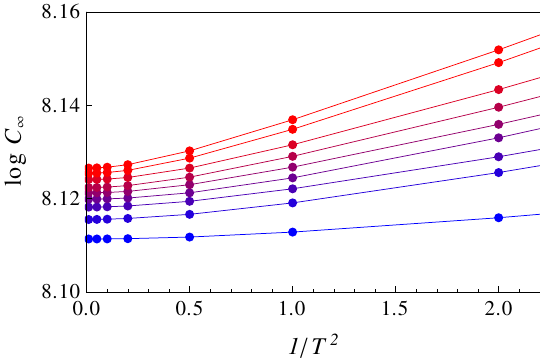}
\caption{\label{fignumotoctemp1} (A) The logarithm of the amplitude of the plateau $C_\infty$ of the kinematic OTOC for the deviations $\sigma=0.05$, $0.1$, $0.2$, $0.4$, $0.6$, $0.8$ (blue to red) as a function of temperature for
$\beta$ values. The linear dependence is nearly perfect, in accordance with the predicted scaling $\log C_\infty\propto 1/T^2$. The matrix size is $N=20$. (B) Same as (A) but for the deviation $\sigma=0.4$ and varying matrix size $N=10$, $20$, $40$, $60$,
$80$, $100$ and $120$ (blue to red). In (C) we bring the zoom-in of the plot (B) for high temperatures. Appart from a slight deviation near $\beta=0$, the behavior for larger matrices is still fully consistent with the analytical prediction. The solid
lines are just to guide the eye.}
\end{figure}

\begin{figure}[ht]
(A)\includegraphics[width=0.4\textwidth]{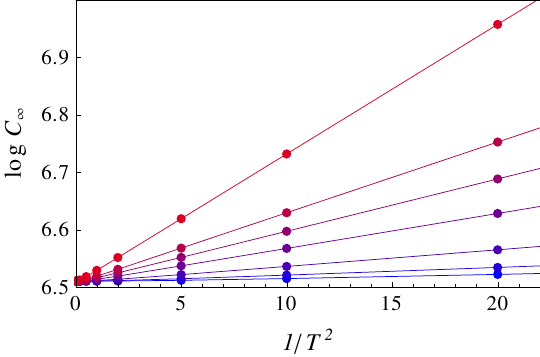}
(B)\includegraphics[width=0.4\textwidth]{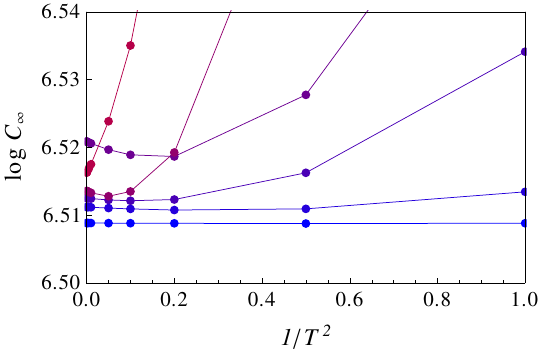}
\caption{\label{fignumotoctemp2} The saturated OTOC $C_\infty(T)$ of a pair of dense random operators $A$ and $B$ for the Gaussian orthogonal random Hamiltonian with $\sigma=0.01$, $0.05$, $0.1$, $0.2$, $0.4$, $0.6$, $0.8$ (blue to red). In (A), looking at
the full range of $C_\infty$ values, it is obvious that the dominant regime is still the $\exp(c/T^2)$ scaling. However, focusing on the low-$\sigma$ ensembles (B), we notice the high-temperature growing behavior of the OTOC plateau which is absent for
sparse operators.}
\end{figure}

Fig.~\ref{fignumotoctemp1} confirms our main prediction for the low-temperature regime (again for the kinematic OTOC) -- clear linear scaling of $\log C_\infty$ with $1/T^2$ in a broad range of temperatures. At small inverse temperatures there is some 
deviation from the linear scaling law but this we also expect. Looking now at the OTOC for a pair of random Hermitian operators in Fig.~\ref{fignumotoctemp2}, we detect also the other regime at small enough temperatures -- $\log C_\infty$ decays with the 
inverse temperature. This regime is likely present also in Fig.~\ref{fignumotoctemp1}, but only at extremely high temperatures (which we have not computed in that figure).

\section{OTOC for weakly and strongly chaotic Hamiltonians}\label{secotocham}

For quasi-integrable few-degrees-of-freedom Hamiltonians one would naively expect that OTOC closely resembles the Lyapunov exponent, at least for high quantum numbers, approaching the classical regime. As we have already commented in the Introduction, it
is known that this is not true in general \cite{lyap,hashimoto17,hashimoto20b,inverse,hashimoto20a} and that both chaotic systems with zero quantum Lyapunov exponent and regular systems with a nonzero exponent exist. We will now try to find some common 
denominator of OTOC dynamics in (deterministic) quantum-mechanical systems. It will quickly become clear from our general analysis of the master formula (\ref{otocmn}) that the function $C(t)$ is as complicated as for random matrices (indeed, even more
so). But we will again construct an upper bound for the saturated OTOC value and arrive at a rough scaling estimate.

\subsection{Weak chaos: perturbation theory}

As an example of a quasi-integrable system (of the form $H=H_0+\epsilon V$ where $H_0$ is integrable and the perturbation $V$ makes it nonintegrable for $\epsilon\neq 0$) consider the Henon-Heiles Hamiltonian
\be
\label{hhham}H=\frac{1}{2}\left(p_x^2+p_y^2\right)+\frac{1}{2}\left(\omega_x^2x^2+\omega_y^2y^2\right)+\epsilon\left(x^2y-\frac{1}{3}y^3\right),
\ee
a well-known paradigm for classical chaos with applications in galactic dynamics and condensed matter. For $\epsilon=0$ it obviously reduces to a 2D linear harmonic oscillator and becomes integrable. As we know, nonintegrability does not always imply
chaos; indeed, this is a typical system with mixed phase space, with both chaotic and regular orbits. Chaotic orbits proliferate only when the perturbation is larger than some finite $\epsilon_c$; they are almost absent at low energies, numerous at
intermediate energies and again absent at very high energies when the potential energy is negligible compared to the kinetic energy \cite{hh1,hh2}. For such a quasi-integrable system our idea is to apply elementary perturbation theory in the occupation
number basis to estimate OTOC starting from (\ref{otocmaster}). We will present the perturbation theory in a fully general way, for an arbitrary Hamiltonian $H_0+\epsilon V$, and some of the conclusions will also turn out to be quite general. Only at
the end we will show the quantitative results for the Henon-Heiles system (\ref{hhham}).

Let us write the perturbative expression for OTOC. Replacing the initial basis states $\vert n\rangle$ with the first-order\footnote{The whole argument works the same way also for higher-order perturbation theory; we assume first order just for
simplicity.} perturbatively corrected states $\vert n+\delta n\rangle$ and introducing likewise the perturbative corrections $\delta a_{mn}$, $\delta b_{mn}$ for the matrix elements of $A$ and $B$, the equation (\ref{otocmn}) becomes
\bea
\nonumber &&c_{mn}^{(1)}=c_{mn}+\sum_k\left(\delta a_{mk}b_{kn}+a_{mk}\delta b_{kn}\right)e^{-\imath E_{km}t}-\sum_k\left(\delta b_{mk}a_{kn}+b_{mk}\delta a_{kn}\right)e^{-\imath E_{nk}t}=\\
\nonumber&&=c_{mn}+\sum_{kl}\left(\delta_{ml}a_{lk}b_{kn}+\delta^*_{kl}a^*_{lm}b_{kn}+\delta_{kl}b_{ln}a_{mk}+\delta^*_{nl}b^*_{lk}a_{mk}\right)e^{-\imath E_{km}t}+\left(a\leftrightarrow b\right)e^{-\imath E_{nk}t}=\\
\label{otocmn1}&&c_{mn}+\left(\delta\cdot A\cdot B+A^\dagger\cdot\delta^\dagger\cdot B+A\cdot\delta\cdot B+A\cdot B^\dagger\cdot\delta^\dagger\right)_{mn}e^{-\imath E_{km}t}+\left(A\leftrightarrow B\right)_{mn}e^{-\imath E_{nk}t}.~~~~~~~~~
\eea
Now we insert this result into (\ref{otocdef1}) and apply the Cauchy-Schwarz-Bunyakovski inequality:
\bea
\nonumber&&C^{(1)}(t)=\frac{1}{Z}\sum_{mn}e^{-\beta E_n}\vert c_{mn}^{(1)}\vert^2\leq\\
\nonumber&&\leq\frac{1}{Z}\sum_{mn}e^{-\beta E_n}\left(\vert c_{mn}\vert^2+\vert\delta\cdot A\cdot B+A^\dagger\cdot\delta^\dagger\cdot B+A\cdot\delta\cdot B+A\cdot B^\dagger\cdot\delta^\dagger\vert_{mn}^2+\left(A\leftrightarrow B\right)\right)=\\
\nonumber&&=C(t)+
\frac{1}{Z}\left(4\mathrm{Tr}\left(B^\dagger\cdot A^\dagger\cdot\delta^\dagger\cdot\tilde{\rho}^2\cdot\delta\cdot A\cdot B\right)+4\mathrm{Tr}\left(B^\dagger\cdot\delta\cdot A\cdot\tilde{\rho}^2\cdot A^\dagger\cdot\delta^\dagger\cdot B\right)\right)\leq\\
\label{cmnsq}&&\leq C_\infty+\frac{8}{Z}\vert\tilde{\rho}\vert^2\vert A\vert^2\vert B\vert^2\vert\delta\vert^2\equiv C_\infty+\delta C_\infty,
\eea
where $\tilde{\rho}\equiv\mathrm{diag}(\exp(-\beta E_n))$ is the non-normalized density matrix. In the above derivation, we have also used the definition of trace and the definition of thermal expectation values
$\langle A\rangle\equiv\mathrm{Tr}\left(\rho\cdot A\right)$. The norm of a matrix is defined as usual by $\vert A\vert^2\equiv\mathrm{Tr}A^\dagger A$. This estimate manifestly replaces $C(t)$ by its asymptotic (maximum) value, as we have replaced the terms
containing the time-dependent phase factors by their time-independent norms.

\begin{figure}[ht]
(A)\includegraphics[width=0.4\textwidth]{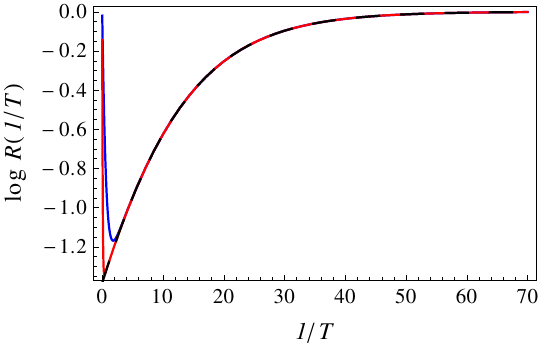}
(B)\includegraphics[width=0.4\textwidth]{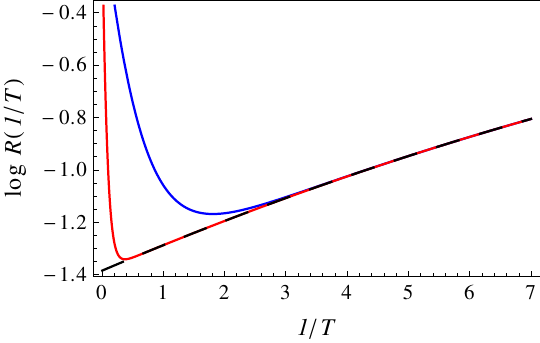}
\caption{\label{figratio} The thermal dependence factor of $\log C_\infty$ for weak perturbative chaotic systems, given by $Z(2\beta)/Z(\beta)$ with $\beta=1/T$ the inverse temperature, here given for a 2D linear harmonic oscillator with the frequencies
$\omega_x=\omega_y=0.1$, with $N=20$ levels (blue) and with $N=150$ levels (red). We also plot the sum over $N=\infty$ levels (black). In (A) we zoom in at high $\beta$/low temperatures, and in (B) we focus on smaller $\beta$/larger $T$. The $N=150$
plot is aready quite close to the monotonic $N=\infty$ dependence but at high temperatures there is always a region decaying with $\beta$, before the approximately linear $\log R(1/T)$ dependence sets in, just like in the numerical results. At very low
temperatures the ratio saturates, as we see in the panel (A). This ensures that our estimate for the saturated OTOC has a finite limite at zero temperature. The overall scale is arbitrary as the R factor is always multiplied by various other factors.}
\end{figure}

In order to move further, notice first that $\vert\tilde{\rho}\vert^2=\sum_n\exp(-2\beta E_n)=Z(2\beta)$ for a canonical ensemble with the diagonal density matrix that we consider here. This means, from (\ref{cmnsq}), that the temperature dependence is
encapsulated in the ratio $Z(2\beta)/Z(\beta)$. The prefactor will again differ between dense and sparse $A$, $B$ and $V$. For sparse matrices, we can write $\vert A\vert^2\sim Na^2$ whereas for dense matrices we have $\vert A\vert^2\sim N^2a^2$, assuming
that all matrix elements have some characteristic magnitude $a$. Obviously, if this is not true the outcome will be more complicated, but it seems this does not influence the temperature dependence. For concreteness we assume sparse $A$ and $B$. For sparse
$V$ with nonzero elements of order $v$ concentrated near the diagonal (this is true for the Henon-Heiles Hamiltonian and in general for perturbations expressed as low-degree polynomials in coordinates and momenta), we can estimate
$\vert\delta\vert^2\sim Nv^2/\omega^2$. Here we assume an approximately equidistant spectrum of $H_0$ with frequency (neighboring level spacing) $\omega$. This yields:
\be
C^{(1)}_\infty\sim C_\infty+\frac{Z(2\beta)}{Z(\beta)}N^3a^2b^2\frac{v^2}{\omega^2}.\label{otocpert}
\ee
For a dense perturbation $V$, the only factor that changes is $\vert\delta\vert$. Assuming again the utterly simple situation where all matrix elements of $V$ are roughly equal $v$, we have $\delta_{mn}\sim v/E_{mn}\sim v/(\omega(m-n))$, which yields a
series that can be summed analytically. However, we will not pursue this further as the temperature dependence is universal in all cases, given by the simple ratio of the partition functions:
\be
\label{otocuni}C_\infty\propto R(\beta)\equiv\frac{Z(2\beta)}{Z(\beta)}\rightarrow\frac{\sum_{j=1}^Ne^{-2\beta\omega}}{\sum_{j=1}^Ne^{-\beta\omega}}\rightarrow\frac{e^{\beta\omega}}{1+e^{\beta\omega}}.
\ee
The first simplification holds when $H_0$ is a 1D harmonic oscillator, and the second one when $N\to\infty$. But the basic result (the ratio of partition functions) always holds. We are in fact more interested in the 2D harmonic oscillator, which is the 
integrable part of the Henon-Heiles Hamiltonian. For that case, we plot the sum (for finite $N$) in Fig.~\ref{figratio}. Of course, the analysis of the function $R(\beta)$ is trivial -- we plot it in the figure merely to emphasize the qualitative agreement
with the numerics. 

As a final remark, what we have found is the correction of the OTOC plateau $\delta C_\infty$. There is still the unperturbed value of $C_\infty$ for the integrable Hamiltonian $H_0$. We know that this can be nonzero and even quite large because of local
instability \cite{lyap,lyapcao,hashimoto20b}. We are mainly interested in the opposite situation, when the scrambling chiefly comes from chaos so that OTOC does not grow when $H=H_0$. In this case $C^{(1)}\approx\delta C_\infty$ and the temperature
dependence is primarily determined by (\ref{otocpert}). In the next subsection we will look both at the Henon-Heiles system where this holds, and a perturbed inverse chaotic oscillator where $H_0$ is unstable.

\subsection{Weak chaos: examples and numerics}

As our main example we can now study the Henon-Heiles system of Eq.~(\ref{hhham}). Starting from the nonperturbed Hamiltonian (2D harmonic oscillator), we will express the nonzero elements of $c_{mn}(t)$ exactly, i.e. we will not use the estimates
(\ref{cmnsq}) as the perturbation is quite simple and amenable to analytic treatment. Denoting a basis state by the quantum number $n=(n_x,n_y)$, we can write the amplitudes $c_{n_xn_yn'_xn'_y}$ as products of amplitudes of the two decoupled subsystems
$c_{n_xn_yn'_xn'_y}=C_{n_xn'_x}C_{n_yn'_y}$, with
\bea
\nonumber C_{n_xn_x}&=&-\imath n_x\omega_x\cos t\\
\nonumber C_{n_x,n_x-2}&=&\frac{\imath}{2}\sqrt{n_x-1}\left(\sqrt{n_x+1}e^{-\imath\omega_xt}-\sqrt{n_x+2}e^{\imath\omega_xt}\right)\\
\label{lhocmn}C_{n_x,n_x+2}&=&\frac{\imath}{2}\sqrt{n_x+1}\left(\sqrt{n_x-1}e^{\imath\omega_xt}-\sqrt{n_x-2}e^{-\imath\omega_xt}\right),
\eea
and all other elements are zero; for the $y$ quantum numbers the coefficients are the same with $(n_x,\omega_x)\mapsto (n_y,\omega_y)$. For nonzero $\epsilon$, the off-diagonal matrix elements can be represented exactly as
\be
c_{n_xn'_xn_yn'_y}(t)=\epsilon\delta_{\vert n_x-n'_x\vert-2}\delta_{\vert n_y-n'_y\vert-1}\sqrt{m_x(m_x-1)}\sqrt{m_y+1},~~m_{x,y}\equiv \mathrm{min}(n_{x,y},n'_{x,y}).\label{lhocmn2}
\ee
The state vectors are now easily calculated in textbook perturbation theory. We have compared the analytic calculation to the numerics and find that they agree 
within a relative error $\leq 0.04$; therefore, one may safely use (\ref{lhocmn}-\ref{lhocmn2}) in order to speed up the computations and avoid numerical diagonalization of large matrices.

\begin{figure}[ht]
(A)\includegraphics[width=0.28\textwidth]{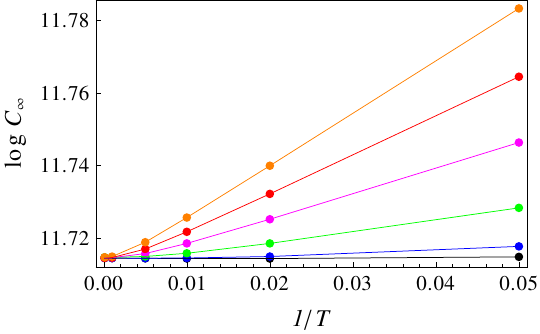}
(B)\includegraphics[width=0.28\textwidth]{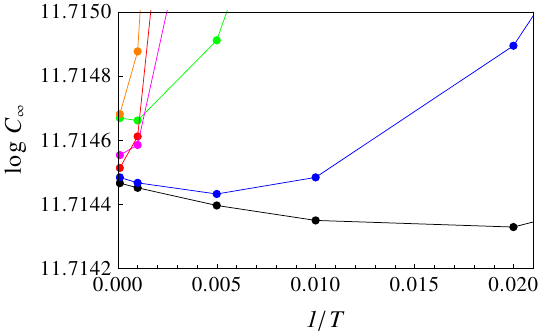}
(C)\includegraphics[width=0.28\textwidth]{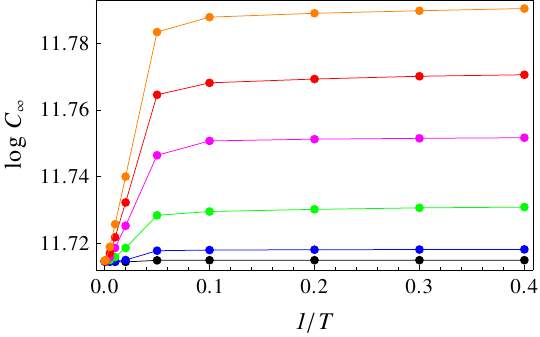}
\caption{\label{fignumhhtemp} (A) The saturated kinematic OTOC value $C_\infty$ for a range of inverse temperatures $\beta=10^{-4}$, $10^{-3}$, $5\times 10^{-3}$, $0.01$, $0.02$, $0.05$ and a range of perturbation strengths $\epsilon=1$, $2$, $5$, $10$,
$15$, $20$ (black, blue, green, magenta, red, orange). Here we see the scaling $C_\infty\propto\exp(c/T)$, with $c$ growing with $\epsilon$. In the (B) panel we zoom in the high-temperature region, to show that for $\epsilon\leq 5$ there is also the other
regime where $C_\infty$ grows with $T$. Since the number of points in this interval is small it is not easy to judge the form of $T$-dependence. In (C) we focus on the opposite regime, at very low temperature, showing that $C_\infty$ saturates as $T\to 0$.
This is again in accordance with the $\beta\to\infty$ limit of $Z(2\beta)/Z(\beta)$. The system is truncated to $N=144$ levels.}
\end{figure}

\begin{figure}[ht]
(A)\includegraphics[width=0.4\textwidth]{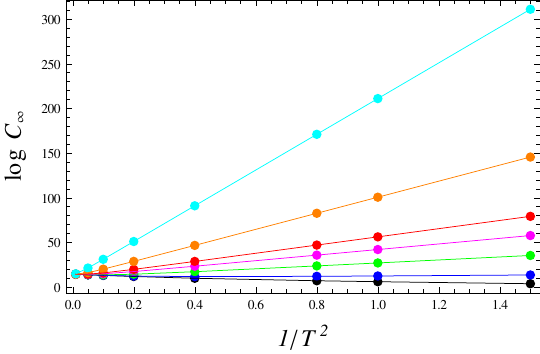}
(B)\includegraphics[width=0.4\textwidth]{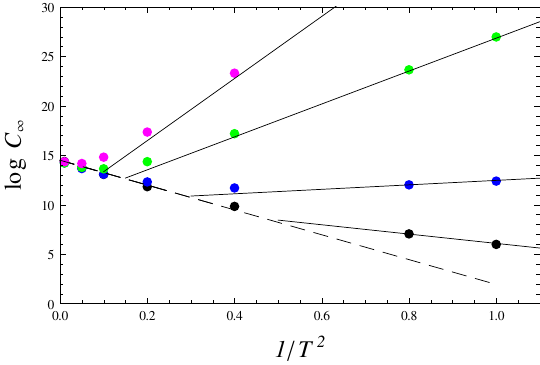}
\caption{\label{fignumhhsmall} The saturated kinematic OTOC value $C_\infty(T)$ for the truncated Henon-Heiles model with $N=25$ levels. The perturbation strength is $\epsilon=0.1$, $0.3$, $0.5$, $0.7$, $0.9$,
$1.5$, $3.0$ (black, blue, green, magenta, red, orange, cyan). Already in (A) we see that for $\epsilon\leq 0.5$ there is a finite crossover temperature $T_c$ so that $C_\infty$ grows wth temperature for $T>T_c$. Since $T_c$ goes down when the Hilbert
space is reduced, we can observe the high-temperature regime very clearly and see that it collapses to a universal law $C_\infty\sim\exp(-4\pi/T)$. This is seen in the panel (B) where we zoom in at the interesting region.}
\end{figure}

The magnitude of the plateau value of $C(t)$,
computed by long-time averaging similarly to the random matrix calculations in Section \ref{secotocrm}, are given in Fig.~\ref{fignumhhtemp}. At large $T$ values, $C_\infty$ decays with $1/T$, at intermediate values it shows an exponential growth with
$1/T$ just like $R(1/T)$ in Fig.~\ref{figratio}, and as the temperature goes to zero it reaches a finite value. In Fig.~\ref{fignumhhsmall} we consider a system with much reduced state space, with $N=25$. We expect that for small $N$ the existence of two
regimes is more clearly visible, and that the crossover temperature is higher. This is indeed what happens, although the exact form of the function $C_\infty(1/T)$ is not very well described by the analytical result. As we have made many crude 
approximations, this is not surprising: our analytical result still explains the existence of two regimes and the crossover between them.\footnote{One might regard such truncation of the state space as artificial and unphysical. It is clearly just a 
technical step in order to show the effect we seek for more clearly, however in principle it can be realized by introducing an additional external potential. In other contexts, e.g. finite spin systems, a finite Hilbert space is perfectly natural.} One
unexpected finding is that the high-temperature regime is apparently universal for all perturbation strengths and scales as $C_\infty\propto\exp(-4\pi/T)$. This is probably specific for the Henon-Heiles system; we do not understand it at present.

\begin{figure}[ht]
(A)\includegraphics[width=0.4\textwidth]{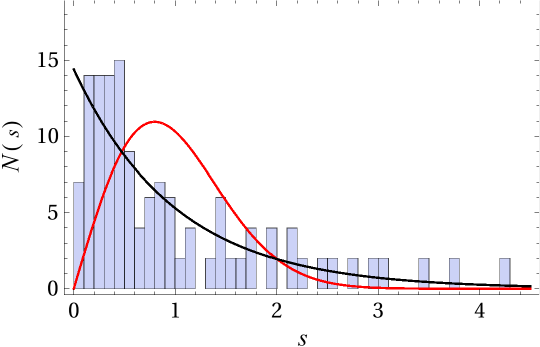}
(B)\includegraphics[width=0.4\textwidth]{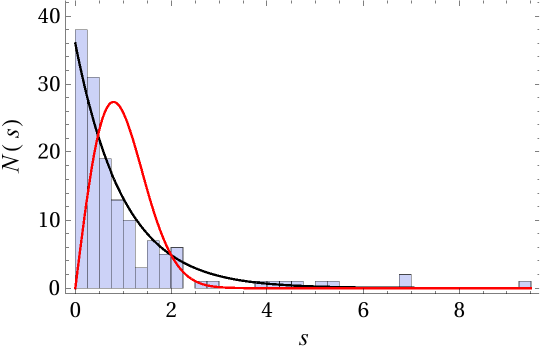}
\caption{\label{figlevels} Distribution of neighboring energy level spacings $N(s)$ for the Henon-Heiles Hamiltonian with $\epsilon=0.1$ (A) and $\epsilon=1.5$ (B). In each plot we compare the level distribution to the Poisson law ($\exp(-s)$) and the 
Wigner-Dyson law for orthogonal matrices ($s\exp(-\pi s^2/4)$). The distribution is dominantly Poissonian even for large perturbations, although there is a small admixture of Wignerian statistics. The perturbative dynamics of the
Henon-Heiles system is is always weakly chaotic in quantum mechanics (despite being classically strongly chaotic for large enough $\epsilon$).}
\end{figure}

It is instructive to look at the energy level statistics of the Henon-Heiles system for the same parameters that we have used for the OTOC calculation, in order to understand the relation of OTOC to chaos. In Fig.~\ref{figlevels} we plot the histograms of
the neighboring level spacing. Even for large $\epsilon$, the regular (Poisson) component is dominant over the chaotic (Wigner-Dyson) component. In other words, the
classically mixed phase space, with the increasing chaotic component, is almost completely regular in the quantum regime; quantum chaos is "weaker", as is often found in the literature \cite{bmnchaos}. For us, the fact that the system is outside the
Wigner-Dyson regime means that indeed the behavior of $C_\infty$ is a good litmus test of quantum dynamics, behaving (at low temperatures) as $\exp(1/T^2)$ or $\exp(1/T)$ for strong or weak chaos respectively.\footnote{In fact, this is only true provided
that the scrambling is chaos-related, i.e. that the integrable limit with $\epsilon=0$ and $H=H_0$ does not scramble significantly. We will come to this issue in the next paragraph.} Indeed, we would not expect that a system which is well
described by perturbation theory shows strong level repulsion.

Finally, it is instructive to look at the inverse Henon-Heiles system, with $(\omega_x^2,\omega_y^2)\mapsto (-\omega_x^2,-\omega_y^2)$, so that $H_0$ is the inverse harmonic oscillator. As already found in the literature, scrambling is significant already
at $\epsilon=0$, and this contribution dominates even at high $\epsilon$, at all temperatures. In other words, neither the perturbation nor the temperature have a significant influence over $C_\infty$. This is fully in accordance with the result
(\ref{cmnsq}) and the morale is that OTOC directly describes scrambling, and chaos only indirectly, through the scrambling, \emph{if the scrambling originates mainly from chaos}; if not, OTOC is largely insensitive to chaos. Therefore, the temperature
dependence of the OTOC value, derived from the assumptions about the dynamics (perturbative chaos or strong, random-matrix chaos) cannot be seen when there is a strong non-chaotic component to scrambling.

\begin{figure}[ht]
\begin{center}\includegraphics[width=0.5\textwidth]{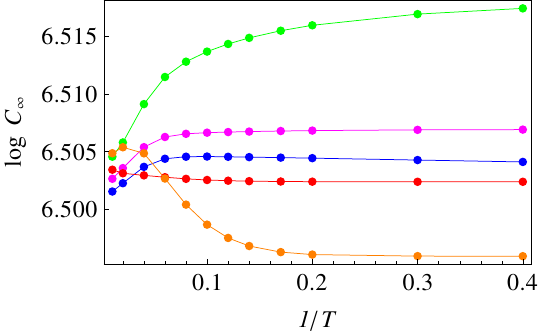}\end{center}
\caption{\label{fignumhhinv} Temperature dependence of the asymptotic OTOC $C_\infty$ for the inverse Henon-Heiles Hamiltonian, with $\omega_x^2=4\omega_y^2=-1$, and perturbation strength $\epsilon=0$, $0.1$, $0.5$, $0.9$, $1.5$ (blue, green, magenta, red,
orange). The curves are more or less flat and without a clear trend, as the scrambling is rooted in local instability, not chaos.}
\end{figure}

\subsection{Strong chaos: numerics and the return to random matrices}

\begin{figure}[ht]
(A)\includegraphics[width=0.28\textwidth]{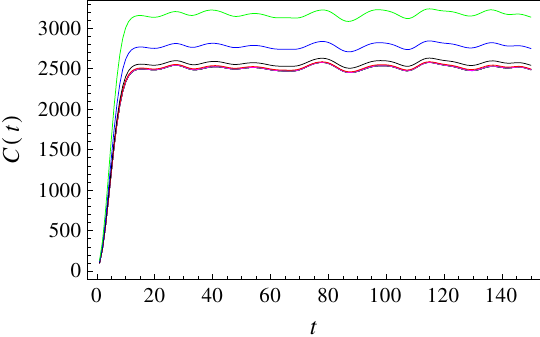}
(B)\includegraphics[width=0.28\textwidth]{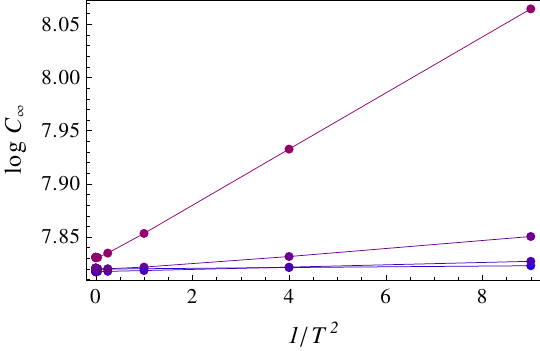}
(C)\includegraphics[width=0.28\textwidth]{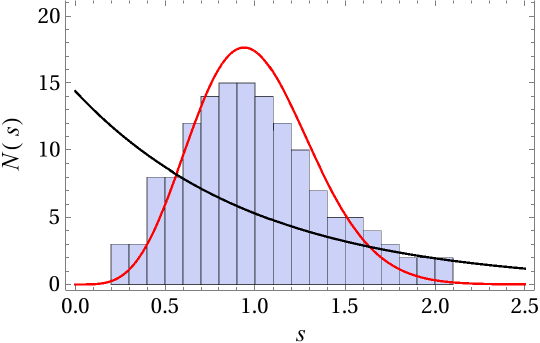}
\caption{\label{figbmn} (A) Time dynamics $C(t)$ for the truncated quantum-mechanical mini-BMN model, with $\nu=0.1$, $0.3$, $0.5$, $1.0$ (red, black, blue, red), shows the expected pattern of growth followed by an oscillating plateau. (B) Temperature
dependence of the saturated value $C_\infty(T)$ for the same values of $\nu$ (blue to red) has the same $\exp(1/T^2)$ scaling as the random matrix ensembles. The level spacing statistics, shown in (C) for $\nu=0.5$, is indeed quite close to the Gaussian
unitary ensemble (full red curve) and clearly at odds with the Poisson statistics (full black curve), confirming that this system is within the scope of our random matrix calculation.}
\end{figure}

As a final stroke, we will now examine a strongly chaotic system which is also relevant for black hole scrambling and related problems in high energy physics. This is the bosonic sector of the D0 brane matrix model known as the BMN
(Berenstein-Maldacena-Nastase) model \cite{bmnmodel}, obtained as a deformation of the BFSS (Banks-Fischler-Shenker-Susskind) model \cite{bfssmodel} by a mass term and a Chern-Simons term. This model has been studied in detail in the context of
non-perturbative string and M theory. It is known to describe the dynamics of M theory on pp-waves and is also related to the type IIA string theory at high energies; for details one can look at the original papers or the review \cite{uspekhi}. Following
\cite{bmnchaos,d0branechaos,hartnollmat,bfssbuyvid}, we focus solely on the bosonic sector which is enough to have strongly chaotic dynamics with equations of motion that are not too complicated. The quantum-mechanical Hamiltonian of the BMN bosonic sector
reads:\footnote{Do not confuse with the classical action considered in \cite{bmnchaos,hartnollmat}.}
\bea
\nonumber H&=&\mathrm{Tr}\left(\frac{1}{2}\Pi^i\Pi^i-\frac{1}{4}\left[X^i,X^j\right]\left[X^i,X^j\right]+\frac{1}{2}\nu^2X^aX^a+\frac{1}{8}\nu^2X^\alpha X^\alpha+\imath\nu\varepsilon_{abc}X^aX^bX^c\right)\\
&&i\in\lbrace 1\ldots 9\rbrace,~~a,b,c\in\lbrace 1,2,3\rbrace,~~\alpha\in\lbrace 4\ldots 9\rbrace,\label{bmnham}
\eea
where $\Pi^i$ are the canonical momenta, $X^i$ the canonical variables, $\varepsilon_{abc}$ is the Levi-Civitta tensor, and $\nu^2>0$ is the mass deformation which also determines the size of the Chern-Simons deformation (the last term in (\ref{bmnham})).
Following \cite{hartnollmat}, we will study the "mini-BMN" model with $X^\alpha=0$, so we effectively only have three degrees of freedom. The matrices $X_{1,2,3}$ and $P_{1,2,3}$ are $N\times N$
matrices. For this example we have to abandon the master formulas for OTOC (\ref{otocdef1}-\ref{otocmn}) as it is very difficult to find the quantities $c_{mn}$ -- for this we would have to perform exact diagonalization of the Hamiltonian (\ref{bmnham}).
Instead, we follow \cite{bfssbuyvid} and write a truncated system of equations directly for the expectation values $\langle X^a\rangle$ and $\langle P^a\rangle$ and the two-point correlators $\langle X^aX^b\rangle$, $\langle \Pi^aX^b\rangle$ and 
$\langle\Pi^a\Pi^b\rangle$, where the expectation value is obtained through the trace over the density matrix: $\langle X^a\rangle\equiv\mathrm{Tr}(\rho X^a)$. The equations read (for their derivation see \cite{bfssbuyvid}):
\bea
\partial_t\langle X^a\rangle&=&\frac{1}{N}\langle\Pi^a\rangle\nonumber\\
\frac{1}{N}\partial_t\langle\Pi^a\rangle&=&
\langle X^b\rangle\langle X^b\rangle\langle X^a\rangle-2\langle X^b\rangle\langle X^a\rangle\langle X^b\rangle+\langle X^a\rangle\langle X^b\rangle\langle X^b\rangle+\nu^2\langle X^a\rangle+
\imath\nu\varepsilon_{abc}\langle X^b\rangle\langle X^c\rangle+\nonumber\\
&&+\left(X^a\langle X^bX^b\rangle-\langle X^bX^b\rangle X^a+X^b\langle X^aX^b\rangle-\langle X^aX^b\rangle X^b+\imath\nu\varepsilon_{abc}\langle X^bX^c\rangle\right),~~~~~~~~~\label{bmneoms}
\eea
where the last line contains the leading quantum corrections: all possible terms with a single contraction of the classical equation of motion, and the summation over repeated indices is understood. The equations of motion for the two-point correlators are
obtained again by writing the classical equations of motion for the bilinears $\Pi^a\Pi^b$, $\Pi^aX^b$ and $X^aX^b$ and taking all possible single contractions in each term. This yields:
\bea
\partial_t\langle X^aX^b\rangle&=&\frac{1}{N}\left(\langle\Pi^aX^b\rangle+\langle X^a\Pi^b\rangle\right)\\
\partial_t\langle\Pi^aX^b\rangle&=&\frac{1}{N}\langle\Pi^a\Pi^b\rangle+N\langle X^aX^b\rangle\langle X^cX^c\rangle-N\langle X^cX^c\rangle\langle X^aX^b\rangle+N\langle X^bX^c\rangle\langle X^aX^c\rangle-\nonumber\\
&&-N\langle X^aX^c\rangle\langle X^bX^c\rangle+\nu^2\langle X^aX^b\rangle\\
\partial_t\langle\Pi^a\Pi^b\rangle&=&N\langle X^aX^b\rangle\langle X^cX^c\rangle-N\langle X^cX^c\rangle\langle X^aX^b\rangle+N\langle X^bX^c\rangle\langle X^aX^c\rangle-\nonumber\\
&&-N\langle X^aX^c\rangle\langle X^bX^c\rangle+\nu^2\langle X^aX^b\rangle+\left(a\leftrightarrow b\right).
\eea
As explained in \cite{bfssbuyvid}, this truncated system is obtained by assuming a Gaussian approximation for the wavefunctions. Therefore, we solve the truncated quantum dynamics of the mini-BMN model -- essentially a toy model, but it will serve our
purpose. Now that we have set the stage, we can express the kinematic OTOC as $C(t)=\langle X^a\Pi^b\rangle-\langle\Pi^aX^b\rangle$ and study its dynamics. The outcome is given in Fig.~\ref{figbmn}. We are essentially back to the random matrix regime of
Section \ref{secotocrm} -- there is a clear scaling $C_\infty\sim\exp(1/T^2)$ (we do not see the other regime, but again it may well be there for sufficiently high temperatures), and the level distribution is a near-perfect fit to the Wigner-Dyson curve.
Therefore, if a Hamiltonian is strongly chaotic, then both the level distribution and the OTOC plateau are well described by the random matrix theory.

\section{Discussion and conclusions}\label{secconc}

In this paper we have formulated a somewhat unexpected indicator of quantum chaos, useful mainly in few-body (few-degrees-of-freedom) systems. While OTOC has become the quintessential object in the studies of quantum chaos and information transport, 
characterized mainly by its growth rate -- the (quantum) Lyapunov exponent, in our examples its growth pattern tends to be quite nonuniversal and "noisy" (in the sense that it depends sensitively on the system at hand and the operators
we look at). Our analytic treatment of OTOC dynamics is quite sketchy, however both analytical and numerical results strongly suggest there is no clear exponential growth. At first glance, one might think that this finding is completely at odds with the
established wisdom, however this is not true. In the literature, exponential growth is mainly characteristic for systems with a classical gravity dual (and reaches its maximum when the dual contains a thermal black hole horizon). There are abundant 
examples of quantum chaotic systems which do not have an exponentially growing OTOC (we especially like \cite{prosenspin} but there are many other published examples). The exponential growth follows, in the AdS/CFT picture, from the shock wave dynamics 
in a classical gravity background, and need not exist when the background is not classical or when the gravity dual does not exist at all. This is precisely what happens here: the Henon-Heiles Hamiltonian is certainly nothing like a strongly coupled large
$N$ field theory, while the truncated mini-BMN model comes closer (it is actually related to discretized Yang-Mills) but we tackle it at finite $N$ and thus away from the fast scrambling dual. For random matrices, our findings for $C(t)$ are in line with
the rigorous results of \cite{cotlerotoc}. As pointed out in that work, the crucial difference between random matrices and strongly coupled field theories is that the former have no notion of locality neither in time nor in space. In our small systems, 
the spatial locality is irrelevant anyway but if the system is not sufficiently chaotic there will still be long-term temporal correlations in dynamics (this indeed gives rise to different scaling regimes for strong and weak chaos).

On the other hand, what we have found is that the long-time OTOC behavior, when it becomes essentially stationary, with a complex oscillation pattern, is surprisingly regular -- behaving as $\exp(1/T^2)$ and $\exp(1/T)$ respectively in strong and weak
chaos. This indicator seems to have a stronger connection to quantum chaos in the sense of level statistics than the Lyapunov exponent; in all examples we have studied the $\exp(1/T^2)$ regime and the Wigner-Dyson level distribution go hand
in hand. At very high temperatures we detect also a different regime, when the OTOC plateau grows with temperature. This regime seems less universal, and we do not understand it very well. One might think that the plateau value should not carry any useful
information; it is often laconically stated that OTOC reaches saturation when the initial perturbation has spread all over the system and that this saturation value is unity when OTOC is appropriately normalized. This is roughly true, however "spreading
all over the system" is not a rigorous notion -- depending on the system and the operators $A$, $B$ in OTOC, the perturbation may never spread completely due to symmetry constraints, specific initial conditions, quasi-integrals of motion etc. Such factors
are particularly important in finite systems (quantum mechanics as opposed to quantum field theory) that we study. Looking at the figures, one sees that differences in the asymptotic OTOC value $C_\infty$ tend to be small, and $C_\infty$ tends to be about
the same to an order of magnitude in all cases. We conjecture that such differences would dwindle to zero in the field limit.

A simple intuitive explanation for the falloff of asymptotic OTOC with temperature is the following: we expect that higher temperatures lead to faster information spreading and quicker equilibration. Therefore, it is logical that the plateau value will be 
lower, so that the system needs less time to reach it, i.e. it needs less time to equilibrate.

We note in passing that we have confirmed that scrambling can originate from at least two distinct mechanisms: local instability and chaos, so in the former case the relation of OTOC to chaos is largeley lost. This is a known fact in many examples already 
\cite{lyap,lyapcao,hashimoto20a,hashimoto20b,kickedhussimi} and we emphasize it here merely as a reminder to the reader that the OTOC-chaos connection is really a relation of three elements: OTOC-scrambling-chaos, and if the second link is missing no
attempt should be made to understand chaotic dynamics from OTOC.

We conclude with some speculations. The OTOC plateau value, as we found, is a rather universal function of temperature, and it is essentially a finite-size fluctuation of the correlation function, when the system is small enough that the relative size of
fluctuations does not go to zero. We may then look for universality and the connections to chaotic dynamics in other similar quantities, e.g. the average fluctuation of the expectation value of some operator during thermalization. Such a quantity remains
nonzero also in AdS/CFT at large $N$, and may relate our results to the more familiar fast scrambling, strongly correlated holographic systems.
 
\section*{Acknowledgments}

This work has made use of the excellent Sci-Hub service. Work at the Institute of Physics is funded by the Ministry of Education, Science and Technological Development and by the Science Fund of the Republic of Serbia, under the
Key2SM project (PROMIS program, Grant No. 6066160).

\appendix

\section{Detailed structure and calculation of OTOC for Gaussian ortohogonal ensembles}\label{secappgoe}

In this Appendix we consider the calculation of OTOC for random matrix systems in some more detail, and describe the detailed structure of the correlation function $C(t)$. Let us first denote, for the sake of brevity:
\begin{equation}
    \sum_\mathrm{tot}\equiv\sum_{n,m}\sum_{k, k'}\sum_{i_1,i_2}\sum_{j_1,j_2}\sum_{i'_1,i'_2}\sum_{j'_1,j'_2},~~\mathbf{C}\equiv c_{j_1}^k ...c_{j'_2}^m\nonumber.
\end{equation}
Denote also the products of matrix elements of the operators $A,B$ entering the expression (\ref{otocmaster}) by $\chi_1,\chi_2,\chi_3,\chi_4$. Now the expression for $\langle C(t)\rangle$ can be written as:
\begin{equation}
    \langle C(t)\rangle=\sum_\mathrm{tot}\int P(\{c\})d^{N^2}\{c\}\int\mathcal{P}(\{E\})d^N{E}e^{-\beta E_n}\mathbf{C}\times\left(\chi_1 e^{i(E_{k'}-E_{k})t}+\ldots+\chi_4 e^{i(E_{m}+E_{n}-E_{k'}-E_{k})t}\right)
\end{equation}
As we have noticed in the main text, the integral over $\mathbf{C}$ yields just a numerical constant. Let us therefore evaluate the energy integral $I_1=\int d^N \{E\}\mathcal{P}(\{E\})e^{-\beta E_n}\chi_1 e^{i(E_{k'}-E_{k})t}$. We have:
\begin{equation}
    I_1=\int \int \dots \int dE_n\int dE_k\int dE_{k'}\mathcal{P}(\{E\})e^{-\beta E_n}\times \chi_1 e^{\imath(E_{k'}-E_{k})t}.\label{appint1}
\end{equation}
The absolute values of the differences can be written out in the obvious way:
\begin{equation}
    \prod_{n<m} |E_n-E_m|=\sum_i (-1)^{\pi(i)}E_1^{\alpha_{1,i}}E_2^{\alpha_{2,i}}...E_N^{\alpha_{N,i}},
\end{equation}
where all $\alpha_{i,j}$ are some (positive) integer exponents and $\pi(i)$ is the appropriate sign factor $-1$ or $1$. Therefore, $I_1$ can be reorganized as:
\begin{equation}
    I_1=\sum_j\chi_1 \prod_{i\neq n,k,k'} \int E_i^{\alpha_{i,j}} e^{-E_i^2}dE_i \int dE_n e^{-\beta E_n}e^{-E_n^2} \int dE_k e^{i E_{k'}t}e^{-E_{k'}^2}\int dE_k e^{-iE_k t}e^{-E_k^2}.
\end{equation}
Note that the part $\prod_{n<m}\vert E_n-E_m\vert$, is not essential for the general behavior, since the singular integral $\int E_i^{\alpha_{i,j}} e^{-E_i^2}dE_i$ is either some constant (if $\alpha$ is even), or zero if $\alpha$ is odd. Otherwise for
$i\neq j$
\begin{equation}
    \int \prod_{l<i<l'}E_i e^{-E_j^2}dE_j=\mathrm{const.}\times\prod_{l<i<l'}E_i.
    \end{equation}
Therefore, we only focus on calculating integrals of the form
\be
\int dE_n\int dE_k\int dE_{k'}e^{-\beta E_n}\times e^{-E_n^2-E_k^2-E_{k'}^2}\chi_1 e^{i(E_{k'}-E_{k})t},\label{appint}
\ee
which yields the closed-form expression for the temperature dependence of $I_1$:
\begin{equation}
    I_1\sim \delta_{k,k'}e^{\beta^2/4}+\left(1-\delta_{k,k'}\right)e^{\beta^2/4}e^{-t^2/2}.\label{appi1},
\end{equation}
where $\delta_{k,k'}$ is the Kronecker delta, reminding us that the main contribution comes from the terms with $E_k=E_{k'}$ which generically means $k=k'$. It is clear that a similar calculation holds for the other parts of $\langle C(t)\rangle$. This
produces the temperature scaling found in the main text for random matrices, of the form $\langle C(t)\rangle\sim e^{1/4T^2}$. But the time dependence is more complicated. In order to see this, we look at the structure of the polynomial factors in $I_1$ in
some more detail. We see immediately that $\langle C(t)\rangle$ will also have dependence on $t^{2n},\beta^n$. Start from
\begin{equation}
    \int E_i^{\alpha_i}e^{-iE_i t}e^{-E_i^2}dE_i=e^{-t^2/4}\int (u-it/2)^{\alpha_i}e^{-u^2}du,
\end{equation}
where $E_i=u-it/2$. Let us look at two cases: $\alpha_i$ even and $\alpha_i$ odd. For any $\alpha_i$ the polynomial will have the form:
\begin{equation}
    (u-it/2)^{\alpha_i}=\sum_{j=0}^{\alpha_i} \gamma_j u^j(it/2)^{\alpha_i-j}.\label{appbin}
\end{equation}
Assume first that $\alpha_i$ is even. This means that $j$ and $\alpha_i-j$ are of same parity. For even $j$ the Gaussian integral evaluates to some constant, but we will also have the prefactor of
$(it/2)^{\alpha_i-j}$, for all even $j\leq\alpha_i$. The odd powers ($j$ odd) will disappear because of the symmetric domain of integration. For $\alpha_i$ odd, $j$ and $\alpha_i-j$ will be of different parity so again, only even
$j$ give a nonzero integral. In conclusion, the integral (\ref{appint}) with polynomial prefactors included will have the form:
\begin{equation}
     \int E_i^{\alpha_i}e^{-iE_i t}e^{-E_i^2}dE_i=e^{-t^2/4}Q(t^{2n}),
\end{equation}
where $Q(t^{2n})$ is a real polynomial depending on even powers of $t$, and $2n\leq \alpha_i$. Alternatively, for $\alpha_i$ odd, we get:
\begin{equation}
     \int E_i^{\alpha_i}e^{-iE_i t}e^{-E_i^2}dE_i=\imath e^{-t^2/4}R(t^{2n+1}),
\end{equation}
where $R(t^{2n+1})$ is a real polynomial depending on odd powers of $t$, and $2n+1\leq \alpha_i$. Analogous logic holds for the $\beta$ dependence. Now we look back at $I_1$:
\begin{equation}
    I_1=\mathrm{const.}\int E_n^{\alpha_n}dE_n\int E_k^{\alpha_k}dE_k\int E_{k'}^{\alpha_{k'}}dE_{k'}e^{-\beta E_n}\times e^{-E_n^2-E_k^2-E_{k'}^2}\chi_1 e^{i(E_{k'}-E_{k})t}.
\end{equation}
When we write out the products of energies, we have the following types of monomials in the resulting polynomial:
\begin{enumerate}
    \item $QQU$
    \item $QRU$
    \item $QQV$
    \item $QRV$,
\end{enumerate}
with the prefactor $\delta_{k,k'}e^{-t^2/2}e^{\beta^2/4}$. Here, $Q,R$ are polynomials of $t$ and are $U$/$V$ are polynomials of even/odd powers of $\beta$ respectively. Note however that $QR$ and $RQ$ give the same structure after integration.

The other integral appearing when writing out the master formula for OTOC is
\begin{equation}
    K_n=\int E_n^{\alpha_n}e^{-\beta E_n}e^{-iE_nt}e^{-E_n^2}dE_n.
\end{equation}
According to the same logic as for $I_1$, it is not hard to get the equivalent form of $K_n$ (leaving out the exponentially decaying terms):
\begin{equation}
    K_n=e^{\beta^2/4}e^{\imath\beta t/2}\int \sum_j \gamma_j u^j (\beta/2+it/2)^{\alpha_n-j}e^{-u^2}du.
\end{equation}
Now we will use the fact that OTOC is a real function, as we can see also from the definition (\ref{otocdef1}). Therefore, all imaginary parts must vanish. From this fact we reach a few important conclusions:
\begin{enumerate}
    \item In the structure of $I_1$, the combination $QR$ is impossible, thus we will only have polynomials of $t$ with an even exponent, and no restriction for polynomials of $\beta$ as it is a real integral, and no term has to
    vanish.
    \item In the structure of $K_n$, when we have the factor $\cos(\beta t/2)$, only even powers of $t$ and arbitrary powers of $\beta$ can survive.
    \item In the structure of $K_n$ when we have the factor $\sin(\beta t/2)$, only odd powers of $t$ and arbitrary powers of $\beta$ can survive.
\end{enumerate}

The conclusion of the above analysis gives us a rough idea of what the $\langle C(t)\rangle$ looks like:
\begin{equation}
    \langle C(t)\rangle=e^{\frac{\beta^2}{4}}W_0(\sigma\beta)+e^{\frac{\beta^2}{4}}
    \left(\cos\left(\frac{\beta t}{2}\right)Q\left(t^{2n}\right)W_1(\sigma\beta)+\sin\left(\frac{\beta t}{2}\right)R\left(t^{2n+1}\right)W_2\left(\beta^n\right)\right),\label{appeqsincos}
\end{equation}
where $W_0,W_1,W_2$ are arbitrary polynomials of $\beta$. This is the form found also in the main text, with the exception that in the main text we have rescaled the combination $\beta t$ as $\beta t/\sigma^2$ in order to have a dimensionless expression.

\subsection{The large matrix limit}


In the limit $N \longrightarrow \infty$ we can say more on the structure of OTOC. We can first schematically rewrite (\ref{appeqsincos}) together with any exponentially suppressed corrections as
\begin{equation}
    \langle C(t)\rangle=e^{\frac{\beta^2}{4}}Q\left(t^{2n}\right)W(\sigma\beta)\left(L_1+L_2e^{-t^2/2}\right).\label{appeqfin}
\end{equation}
Here we have first absorbed all time and $\beta$ dependence of (\ref{appeqsincos}) into the functions $Q$ and $W$ respectively, and then we have included the exponentially suppressed correction coming from the $k\neq k'$ terms in the integrals $I_1$ and
$K_n$. By $L_1,L_2$ we denote the constant (time- and temperature-independent) factors. In general one can write $L_1$ as
\begin{equation}
    L_1=\sum_{j=1}^{N}\sum_{i=0}^j c_i \binom{j}{i}
\end{equation}
We can easily estimate the second sum. Namely,
\begin{equation}
    \Bigg(\sum_{i=0}^j c_i \binom{j}{i}\Bigg)^2\leq \Bigg(\sum_{i=0}^j c_i^2 \Bigg)\Bigg(\sum_{i=0}^j \binom{j}{i}^2 \Bigg),
\end{equation}
by the Cauchy-Schwarz-Bunyakovski inequality. Next, the well known formula $\sum_{i=0}^j \binom{j}{i}^2=\binom{2j}{j}$ yields
\begin{equation}
    \sum_{i=0}^j c_i \binom{j}{i}\leq\mathrm{const.}\times\sqrt{\binom{2j}{j}}.
\end{equation}
To get rid of the binomial coefficient we will use the Stirling's formula and get
\begin{equation}
    \sqrt{\binom{2j}{j}}=\sqrt{\frac{(2j)!}{j!j!}}\approx\sqrt{\frac{\sqrt{4\pi j}\frac{(2j)^{2j}}{e^{2j}}}{2\pi j \frac{(j^j)^2}{(e^j)^2}}}\approx\mathrm{const.}\times\frac{2^j}{j^{1/4}}.
\end{equation}
Finally we reach the result:
\begin{equation}
L_1\approx\mathrm{const.}\times\sum_{j=1}^{N}\frac{2^j}{j^{1/4}}\approx\mathrm{const.}\times\frac{2^{N+1}}{N^{1/4}},
\end{equation}
for $N\longrightarrow\infty$. Exactly the same logic goes for $L_2$.

In the large matrix limit it is possible to show explicitly what we know has to happen: OTOC reaches a plateau. Looking at (\ref{appeqfin}), the condition to reach the plateau for times longer than some scale $t_0$ is
\begin{equation}
    e^{\frac{-t^2}{2}}Q(t^{2n})(L_1+L_2e^{-t^2/2})=\mathrm{const.}~~~t>t_0.
\end{equation}
It is more convenient to look at the forms given in (\ref{appeqsincos}). First let us look at the condition $Q(t^{2n})=\mathrm{const.}\times e^{t^2/2}$. The exponential term can be represented as a series; equating it with $Q(t^{2n})$ we get
\begin{equation}
    \sum_j\alpha_jt^{2j}=\mathrm{const.}\times\sum_j \frac{t^{2j}}{2^j j!},
\end{equation}
thus, we need $\alpha_j \sim \frac{1}{2^j j!}$, which we know is the case from (\ref{appbin}). For the second term the situation is similar:
\begin{equation}
    \sum_j \beta_j t^{2j}=\mathrm{const.}\times \sum_j \frac{t^{2j}}{j!},
\end{equation}
so we need to have $\beta_j \sim \frac{1}{j!}$; this is true by $\cos(\beta t/2)=Q(t^{2n})W(\beta^{2n})$ and $\sin(\beta t/2)=Q(t^{2n+1})W(\beta^{2n+1})$, since the terms in the Taylor expansions of the left-hand sides behave as $ \sim \frac{1}{j!}$.


We can also look at the opposite limit in which $t\longrightarrow 0$. Let us rearrange (\ref{appeqfin}):
\begin{equation}
    \langle C(t)\rangle=L_1' Q(t^{2n})e^{-t^2/2}+L_2'Q(t^{2n})e^{-t^2}.
\end{equation}
Now, simply using the definition of $Q$ and expanding into a series we get:
\begin{equation}
    \langle C(t)\rangle=L_1'\left(1-\frac{t^2}{2}+o\left(t^4\right)\right)\left(q_o+q_1t^2\right)+L_2'\left(1-t^2+o\left(t^4\right)\right)\left(q_o+q_1t^2\right).
\end{equation}
After some algebra we get:
\begin{equation}
    \langle C(t)\rangle=Q_0+Q_1t^2+Q_2t^4+o\left(t^4\right)=P(t).
\end{equation}
We see now that OTOC behaves in a very simple way for early times; this expansion is also consistent with the result (\ref{cotler}) of \cite{cotlerotoc}.


\begin{thebibliography}{20}

\bibitem{chaosbnd} J.~Maldacena, S.~H.~Shenker and D.~Stanford, \emph{A bound on chaos}, JHEP~{\bf08} (2016) 106.
[arXiv:1503.01409[hep-th]]
\bibitem{scramblestrings} S.~H.~Shenker and D.~Stanford, \emph{Stringy effects in scrambling}, JHEP~{\bf05} (2015) 132.
[arXiv:1412.6087[hep-th]]
\bibitem{chaosbnds} S.~Kundu, \emph{Extremal chaos}, (2021).
[arXiv:2109.08693[hep-th]]
\bibitem{scramble} J.~Sekino and L.~Susskind, \emph{Fast scramblers}, JHEP~{\bf10} (2008) 065.
[arXiv:0808.2096[hep-th]]
\bibitem{scramble2} N.~Lashkari, D.~Stanford, M.~Hastings, T.~Osborne and P.~Hayd, \emph{Towards the fast scrambling conjecture}, JHEP~{\bf04} (2013) 022.
[arXiv:1111.6580[hep-th]]
\bibitem{butterstring} S.~H.~Shenker and D.~Stanford, \emph{Stringy effects in scrambling}, JHEP~{\bf05} (2015) 132.
[arXiv:1412.6087[hep-th]]
\bibitem{lyap} E.~B.~Rozenbaum, L.~A.~Bunimovich and V.~Galitski, \emph{Early-time exponential instabilities in nonchaotic quantum systems}, Phys.~Rev.~Lett.~{\bf125} 014101 (2020).
[arXiv:1902.05466[quant-ph]]
\bibitem{lyapcao} T.~Xu, T.~Scaffidi and X.~Cao, \emph{Does scrambling equal chaos?}, Phys.~Rev.~Lett.~{\bf124}, 140602 (2020).
[arXiv:1912.11063[cond-mat.stat-mech]]
\bibitem{hashimoto17} K.~Hashimoto, K.~Murata and R.~Yoshii, \emph{Out-of-time-order correlators in quantum mechanics}, JHEP~{\bf10} (2017) (138).
[arXiv:1703.09435[hep-th]]
\bibitem{hashimoto20b} K.~Hashimoto, K.-B.~Huh, K.-Y.~Kim and R.~Watanabe, \emph{Exponential growth of out-of-time-order correlator without chaos: inverted harmonic oscillator}, JHEP~{\bf11} (2020) 068.
[arXiv:2007.04746[hep-th]]

\bibitem{thebook} F.~Haake, S.~Gnutzman, M.~Ku\'s, \emph{Quantum signatures of chaos}, Springer-Verlag, Berlin, 2019.
\bibitem{randmat} M.~L.~Mehta, \emph{Random matrices}, Academic, New York, 2004.

\bibitem{cotlerbig} J.~S.~Cotler et al, \emph{Black holes and random matrices}, JHEP~{\bf05} (2017) 118.
[arXiv:1611.04650[hep-th]]
\bibitem{cotlerotoc} J.~Cotler, N.~Hunter-Jones, J.~Liu and B.~Yoshida, \emph{Chaos, complexity, and random matrices}, JHEP~{\bf11} (2017) 048.
[arXiv:1706.05400[hep-th]]
\bibitem{cotlerdecoup} J.~Cotler and N.~Hunter-Jones, \emph{Spectral decoupling in many-body quantum chaos}, JHEP~{\bf12} (2020) 205.
[arXiv:1911.02026[hep-th]]

\bibitem{false} W.~Kirkby, D.~H.~J.~O'Dell and J.~Mumford, \emph{False signals of chaos from quantum probes}, (2021).
[arXiv:2108.09391[hep-th]]
\bibitem{hashimoto20a} T.~Akutagawa, K.~Hashimoto, T.~Sasaki and R.~Watanabe, \emph{Out-of-time-order correlator in coupled harmonic oscillators}, JHEP~{\bf08} (2020) 013.
[arXiv:2004.04381[hep-th]]
\bibitem{inverse} A.~Bhattacharyya, W.~Chemissany, S.~S.~Haque, J.~Murugan and B.~Yan, \emph{The multi-faceted inverted harmonic oscillator: chaos and complexity}, SciPost~Phys.~Core~{\bf4}, 002 (2021).
[arXiv:2007.01232[hep-th]]
\bibitem{poljaci} W.~Klobus et al, \emph{Transition from order to chaos in reduced quantum dynamics}, (2021).
[arXiv:2111.13477[quant-ph]]
\bibitem{prosenspin} B.~Bertini, P.~Kos and T.~Prosen, \emph{Exact spectral form factor in a minimal model of many-body quantum chaos}, Phys.~Rev.~Lett.~{\bf121} 264101 (2018).
[arXiv:1805.00931[nlin.CD]]

\bibitem{pappal} S.~Pappalardi, A.~Russomanno, B.~\v{Z}unkovi\v{c}, F.~Iemini, A.~Silva and R.~Fazio, \emph{Scrambling and entanglement spreading in long-range spin chains}, Phys.~Rev.~B~{\bf98}, 134303 (2018).
[arXiv:1806.00222[quant-ph]]
\bibitem{okuyama} K.~Okuyama, \emph{Spectral form factor and semi-circle law in the time direction}, JHEP~{\bf02} (2019) 161.
[arXiv:1811.09988[quant-ph]]
\bibitem{richter} Q.~Hummel, B.~Geiger, J.~D.~Urbina and K.~Richter, \emph{Reversible quantum information spreading in many-body systems near criticality}, Phys.~Rev.~Lett.~{\bf123}, 160401 (2019).
[arXiv:1812.09237[quant-ph]]
\bibitem{kerr} H.~Goto and T.~Kanao, \emph{Chaos in coupled Kerr-nonlinear parametric oscillators}, (2021).
[arXiv:2110.04019[quant-ph]]
\bibitem{doublekicked} N.~D.~Varikuti and V.~Madhok, \emph{Out-of-time ordered correlators in kicked coupled tops and the role of conserved quantities in information scrambling}, (2022).
[arXiv:2201.05789[quant-ph]]
\bibitem{rabi} A.~V.~Kirkova, D.~Porras and P.~A.~Ivanov, \emph{Out-of-time-order correlator in the quantum Rabi model}, (2022).
[arXiv:2201.06340[quant-ph]]
\bibitem{kickedhussimi} J.~R.~G.~Alonso, N.~Shammah, S.~Ahmed, F.~Nori and J.~Dressel, \emph{Diagnosing quantum chaos with out-of-time-ordered-correlator quasiprobability in the kicked-top model}, (2022).
[arXiv:2201.08175[quant-ph]]
\bibitem{energylimit} K.~Hashimoto, K.~Murata, N.~Tanahashi and R.~Watanabe, \emph{A bound on energy dependence of chaos}, (2021).
[arXiv:2112.11163[hep-th]]

\bibitem{hh1} J.~Aguirre, J.~C.~Vallejo, M.~A.~F.~Sanju\'an, \emph{Wada basins and chaotic invariant sets in the Henon-Heiles system}, Phys.~Rev.~E{\bf64} 066208 (2001).
\bibitem{hh2} H.~E.~Kandrup, C.~Siopis, G.~Contopoulos and R.~Dvorak, \emph{Diffusion and scaling in escapes from two-degrees-of-freedom Hamiltonian systems}, Chaos~{\bf9} 381 (1999).
[arXiv:astro-ph/9904046]

\bibitem{bmnmodel} D.~E.~Berenstein, J.~M.~Maldacena and H.~S.~Nastase, \emph{Strings in flat space and pp waves from N=4 super Yang-Mills}, JHEP~{\bf04} (2002) 013.
[arXiv:hep-th/0202021]
\bibitem{bfssmodel} T.~Banks, W.~Fischler, S.~H.~Shenker and L.~Susskind, \emph{M theory as a matrix model: a conjecture}, Phys.~Rev~D{\bf55} 5112 (1997).
[arXiv:hep-th/9610043]
\bibitem{uspekhi} K.~L.~Zarembo and Yu.~M.~Makeenko, \emph{An introduction to matrix superstring models}, Physics-Uspekhi~{\bf41}, 1 (1998) (in Russian). 

\bibitem{bmnchaos} Y.~Asano, D.~Kawai and K.~Yoshida, \emph{Chaos in the BMN matrix model}, JHEP~{\bf06} (2015) 191.
[arXiv:1503.04594[hep-th]]
\bibitem{d0branechaos} Guy~Gur-Ari, M.~Hanada and S.~H.~Shenker, \emph{Chaos in classical D0-brane mechanics}, JHEP~{\bf02} (2016) 091.
[arXiv:1512.00019[hep-th]]
\bibitem{hartnollmat} X.~Han and S.~A.~Hartnoll, \emph{Deep quantum geometry of matrices}, Phys.~Rev~X{\bf10} 011069 (2020).
[arXiv:1906.08781[hep-th]]
\bibitem{bfssbuyvid} P.~V.~Buividovich, M.~Hanada and A.~Sch\"{a}fer, \emph{Quantum chaos, thermalization and entanglement generation in real-time simulations of the BFSS matrix model}, Phys.~Rev~D{\bf99} 046011 (2019).
[arXiv:1810.03378[hep-th]]

\end{thebibliography}
\end{document}